\documentclass[12pt]{article}
\pdfoutput=1
\usepackage[totalwidth=480pt,totalheight=640pt]{geometry}
\usepackage[centertags]{amsmath}
\usepackage[square, comma, sort&compress,numbers]{natbib}
\usepackage{array,multirow}
\usepackage{amssymb}
\usepackage{graphicx}
\usepackage{color}
\usepackage{setspace}

\usepackage{epsfig}

\def\Or[#1]{{\text{O}}\left({#1}\right)}
\def\dotl[#1,#2]{\left\langle #1, #2 \right\rangle}
\def\dotlb[#1,#2]{[ #1, #2 ]}
\def\dotp[#1,#2]{(#1) \cdot (#2)}
\def\aff[#1,#2]{\hat{#1}(#2)}
\def\n4sym{{\cal N}=4 SYM}
\def\>{\rangle}
\def\<{\langle}
\def\weight[#1,#2,#3]{\{(#1),#2,#3\}}
\def\ads[#1]{$\text{AdS}_{#1}$}

\newcommand{\ba}{\begin{eqnarray}}
\newcommand{\ea}{\end{eqnarray}}
\newcommand{\be}{\begin{eqnarray}}
\newcommand{\ee}{\end{eqnarray}}

\newcommand{\CL}{{\cal L}}

\newcommand{\CO}{{\cal O}}

\newcommand{\nn}{\nonumber}

\newcommand\oo\infty
\newcommand\s\sigma
\newcommand\de\delta
\newcommand\De\Delta
\newcommand\p[1]{\left(#1\right)}
\newcommand\f\phi
\newcommand\g\gamma
\newcommand\x\times
\usepackage{hyperref}

\begin{document}

\begin{titlepage}

\begin{center}
\vspace{1cm}

{\Large \bf  The Analytic Bootstrap and AdS Superhorizon Locality}

\vspace{0.8cm}

\small
\bf{A. Liam Fitzpatrick$^1$,  Jared Kaplan$^{1,2}$, David Poland$^3$, David Simmons-Duffin$^4$}
\normalsize

\vspace{.5cm}

{\it $^1$ Stanford Institute for Theoretical Physics, Stanford University, Stanford, CA 94305}\\
{\it $^2$ Department of Physics and Astronomy, Johns Hopkins University, Baltimore, MD 21218} \\
{\it $^3$ Department of Physics, Yale University, New Haven, CT 06520} \\
{\it $^4$ School of Natural Sciences, Institute for Advanced Study, Princeton, NJ 08540} \\

\end{center}

\vspace{1cm}

\begin{abstract}

We take an analytic approach to the CFT bootstrap, studying the 4-pt correlators of $d>2$ dimensional CFTs in an Eikonal-type limit, where the conformal cross ratios satisfy $|u| \ll |v| < 1$.  We prove that every CFT with a scalar operator $\phi$ must contain infinite sequences of operators $\CO_{\tau, \ell}$ with twist approaching $\tau \to 2 \Delta_{\phi} + 2n$ for each integer $n$ as $\ell \to \infty$. We show how the rate of approach is controlled by the twist and OPE coefficient of the leading twist operator in the $\phi \times \phi$ OPE, and we discuss SCFTs and the 3d Ising Model as examples. Additionally, we show that the OPE coefficients of other large spin operators appearing in the OPE are bounded as $\ell \to \infty$.  We interpret these results as a statement about superhorizon locality in AdS for general CFTs.

\end{abstract}

\bigskip

\end{titlepage}

\section{Introduction and Review}
\label{sec:Introduction}

The last few years have seen a remarkable resurgence of interest in an old approach to Conformal Field Theory (CFT), the conformal bootstrap \cite{Polyakov:1974gs, Belavin:1984vu}, with a great deal of progress leading to new results of phenomenological \cite{Rattazzi:2008pe, Rychkov:2009ij,  Rattazzi:2010yc, Vichi:2011ux, Poland:2011ey, ElShowk:2012ht, ElShowk:2012hu} and theoretical \cite{JP, Rattazzi:2010gj, Poland:2010wg, Heemskerk:2010ty, Pappadopulo:2012jk, Liendo:2012hy} import.  Most of these new works use numerical methods to constrain the spectrum and OPE coefficients of general CFTs.  In a parallel series of developments, there has been significant progress understanding effective field theory in AdS and its interpretation in CFT \cite{JP, Heemskerk:2010ty, Katz, Sundrum:2011ic, ElShowk:2011ag, AdSfromCFT}.  This has led to a general bottom-up classification of which CFTs have dual \cite{Maldacena, Witten, Gubser:1998bc} descriptions as effective field theories in AdS, providing an understanding of AdS locality on all length scales greater than the inverse energy cutoff in the bulk.   In fact, these two developments are closely related, as the seminal paper \cite{JP} and subsequent work begin by applying the bootstrap to the $1/N$ expansion of CFT correlators.  This approach has been fruitful, especially when interpreted in Mellin space \cite{Mack, MackSummary, JoaoMellin, NaturalLanguage, Paulos:2011ie, Paulos:2012nu, AdSfromCFT}, but it is an essentially perturbative approach analogous \cite{Unitarity} to the use of dispersion relations for the study of perturbative scattering amplitudes.

In light of recent progress, one naturally wonders if an analytic approach to the bootstrap could yield interesting new exact results.  In fact, in \cite{Pappadopulo:2012jk} bounds on operator product expansion (OPE) coefficients for large dimension operators have already been obtained.  We will obtain a different sort of bound on both OPE coefficients and operator dimensions in the limit of large angular momentum, basically providing a non-perturbative bootstrap proof of some results that Alday and Maldacena \cite{Alday:2007mf} have also discussed.\footnote{ The authors of \cite{Alday:2007mf} explicitly discuss minimal twist double-trace operators in a large $N$ gauge theory; however their elegant argument can be applied in a more general context, beyond perturbation theory and for general twists.  We thank J. Maldacena for discussions of this point. }

Specifically, we will study a general scalar primary operator $\phi$ of dimension $\Delta_{\phi}$ in a CFT in $d>2$ dimensions.  We will prove that for each non-negative integer $n$ there must exist an infinite tower of operators $\CO_{\tau, \ell}$ with twist $\tau \to 2 \Delta_{\phi} + 2n$ appearing in the OPE of $\phi$ with itself.  This means that at large $\ell$, and we can define an `anomalous dimension' $\gamma(n, \ell)$ which vanishes as $\ell \to \infty$.  If there exists one such operator at each $n$ and $\ell$, we will argue that at large $\ell$ the anomalous dimensions should roughly approach
\be
\gamma(n, \ell) \approx \frac{\gamma_n}{\ell^{\tau_m}} ,
\ee
where $\tau_m$ is the twist of the minimal twist operator appearing in the OPE of $\phi$ with itself.  Related predictions can be made about the OPE coefficients.  Finally, we will show that the OPE coefficients of other operators appearing in the OPE of $\phi$ with itself at large $\ell$ must be bounded, so that they fall off even faster as $\ell \to \infty$.  Similar results also hold for the OPE of pairs of operators $\phi_1$ and $\phi_2$, although for simplicity we will leave the discussion of this generalization to Appendix \ref{app:DistinctOperators}.

Our arguments fail for CFTs in two dimensions, and in fact minimal models provide an explicit counter-example.  Two dimensional CFTs are distinguished because there is no gap between the twist of the identity operator and the twist of other operators, such as conserved currents and the energy-momentum tensor.  

Our results can be interpreted as a proof that all CFTs in $d > 2$ dimensions have correlators that are dual to local AdS physics on superhorizon scales.  That is, CFT processes that are dual to bulk interactions will effectively shut off as the bulk impact parameter is taken to be much greater than the AdS length.  This can also be viewed as a strong form of the cluster decomposition principle in the bulk.  Since the early days of AdS/CFT it has been argued that this notion of ``coarse locality'' \cite{JP} could be due to a decoupling of modes of very different wavelengths, but it has been challenging to make this qualitative holographic RG intuition precise.  The bootstrap offers a precise and general method for addressing coarse locality.

For the remainder of this section we will give a quick review of the CFT bootstrap.  Then in section~\ref{sec:Proof} we delve into the argument, first giving an illustrative example from mean field theory (a Gaussian CFT, with all correlators fixed by 2-pt functions, e.g. a free field theory in AdS).  We give the complete argument in sections \ref{sec:DblTrace} and \ref{sec:Bound}, with some more specific results and examples that follow from further assumptions in section \ref{sec:IsolatedTowers}. We provide more detail on how two dimensional CFTs escape our conclusions in section \ref{sec:2d}.  In section \ref{sec:super} we connect our results to superhorizon locality in AdS, and we conclude with a brief discussion in section \ref{sec:Discussion}.  In Appendix \ref{app:LargeLBlocks} we collect some results on relevant approximations of the conformal blocks in four and general dimensions.  In Appendix \ref{app:Rigorous} we give a more formal and rigorous version of the argument in section \ref{sec:Proof}.  In Appendix \ref{app:DistinctOperators} we explain how our results generalize to terms occurring in the OPE of distinct operators $\phi_1$ and $\phi_2$.  In Appendix \ref{app:Joao} we connect our results with perturbative gravity computations in AdS.

\emph{Note added: after this work was completed we learned of the related work of Komargodski and Zhiboedov \cite{Komargodski:2012ek}; they obtain very similar results using somewhat different methods. }

\subsection{Lightning Bootstrap Review}

In CFTs, the bootstrap equation follows from the constraints of conformal invariance and crossing symmetry applied to the operator product expansion, which says that a product of local operators is equivalent to a sum
\be
\phi(x) \phi(0) &=& \sum_{\CO} c_{\CO} f_{\CO}(x,\partial) \CO(0).
\ee
  Conformal invariance relates the OPE coefficients of all operators in the same irreducible conformal multiplet, and this allows one to reduce the sum above to a sum over different irreducible multiplets, or ``conformal blocks''.  When this expansion is performed inside of a four-point function, the contribution of each block is just a constant ``conformal block coefficient'' $P_{\CO} \propto c_{\CO}^2$ for the entire multiplet times a function of the $x_i$'s whose functional form depends only on the spin $\ell_{\CO}$ and dimension $\Delta_{\CO}$ of the lowest-weight (i.e. ``primary'') operator of the multiplet:
  \be
  \< \phi(x_1) \phi(x_2) \phi(x_3) \phi(x_4) \> &=&  \frac{1}{(x_{12}^2 x_{34}^2)^{\Delta_{\phi}}} \sum_{\CO} P_{\CO} g_{\tau_{\CO}, \ell_{\CO}}(u,v),
  \ee
where $x_{ij} = x_i - x_j$, the twist of $\CO$ is $\tau_{\CO} \equiv \Delta_{\CO} - \ell_{\CO}$, and 
\be
u= \left( \frac{x_{12}^2x_{34}^2}{x_{24}^2 x_{13}^2 } \right), \qquad  v = \left( \frac{x_{14}^2 x_{23}^2}{x_{24}^2 x_{13}^2 } \right),
\ee
 are the conformally invariant cross-ratios.  The functions $g_{\tau_{\CO}, \ell_{\CO}} (u,v)$ are also usually referred to as conformal blocks or conformal partial waves \cite{Dolan:2000ut, Dolan:2003hv, Dolan:2011dv, DSDProjectors}, and they are crucial elementary ingredients in the bootstrap program.
 
 In the above, we took the OPE of $\phi(x_1) \phi(x_2)$ and $\phi(x_3) \phi(x_4)$ inside the four-point function, but one can also take the OPE in the additional ``channels'' $\phi(x_1) \phi(x_3)$ and $\phi(x_2) \phi(x_4)$ or $\phi(x_1) \phi(x_4)$ and $\phi(x_2) \phi(x_3)$, and the bootstrap equation is the constraint that the decomposition in different channels matches:
 \be
 \label{eq:bootstrap}
   \frac{1}{(x_{12}^2 x_{34}^2)^{\Delta_{\phi}}} \sum_{\CO} P_{\CO}g_{\tau_{\CO}, \ell_{\CO}} (u,v) &=&
     \frac{1}{(x_{14}^2 x_{23}^2)^{\Delta_{\phi}}} \sum_{\CO} P_{\CO}g_{\tau_{\CO}, \ell_{\CO}} (v,u).
\ee
Much of the power of this constraint follows from the fact that by unitarity, the conformal block coefficients $P_{ \CO}$ must all be non-negative in each of these channels, because the $P_{\CO}$ can be taken to be the squares of real OPE coefficients.

\section{The Bootstrap and Large $\ell$ Operators}
\label{sec:Proof}

Although some of the arguments below are technical, the idea behind them is very simple.  By way of analogy, consider the $s$-channel partial wave decomposition of a tree-level scattering amplitude with poles in both the $s$ and $t$ channels.  The center of mass energy is simply $\sqrt{s}$, so the $s$-channel poles will appear explicitly in the partial wave decomposition.  However, the $t$-channel poles will not be manifest.  They will arise from the infinite sum over angular momenta, because the large angular momentum region encodes long-distance effects.  Crossing symmetry will impose constraints  between the $s$-wave and $t$-wave decompositions, relating the large $\ell$ behavior in one channel with the pole structure of the other channel.   

We will be studying an analogous phenomenon in the conformal block (sometimes called conformal partial wave) decompositions of CFT correlation functions.  The metaphor between scattering amplitudes and CFT correlation functions is very direct when the CFT correlators are expressed in Mellin space, but in what follows we will stick to position space. In position space CFT correlators, the poles of the scattering amplitude are analogous to specific power-laws in conformal cross-ratios, with the smallest power-laws corresponding to the leading poles.

\subsection{An Elementary Illustration from Mean Field Theory}
\label{sec:MFTExample}

Let us begin by considering what naively appears to be a paradox.  Consider the 4-point correlation function in a CFT with only Gaussian or `mean field theory' (MFT) type correlators. 
These mean field theories are the dual of free field theories in AdS.  We will study the 4-pt correlator of a dimension $\Delta_{\phi}$ scalar operator $\phi$ in such a theory. 
By definition,  in mean field theory the 4-pt correlator is given as a sum over the 2-pt function contractions:
\be
\langle \phi(x_1) \phi(x_2)  \phi(x_3) \phi(x_4) \rangle &=& \frac{1}{(x_{12}^2 x_{34}^2)^{\Delta_{\phi}} } + \frac{1}{(x_{13}^2 x_{24}^2)^{\Delta_{\phi}}} + \frac{1}{(x_{14}^2x_{23}^2)^{\Delta_{\phi}}} ,\nn\\
&=& \frac{1}{(x_{13}^2 x_{24}^2)^{\Delta_{\phi}}} \left(u^{-\Delta_{\phi}} + 1  + v^{-\Delta_{\phi}} \right).
\ee

Since this is the 4-pt correlator of a unitary CFT, it has a conformal block decomposition in every channel with positive conformal block coefficients.  The operators appearing in the conformal block decomposition are just the identity operator $\mathbf{1}$ and the ``double-trace'' operators $\CO_{n,\ell}$ of the schematic form
\be
\CO_{n,\ell} \sim \phi (\partial^2)^n \partial_{\mu_1} \dots \partial_{\mu_\ell} \phi,
\ee
with known \cite{Unitarity} conformal block coefficients $P_{2\Delta_{\phi} + 2n,\ell}$ and twists $\tau_{n,\ell} = 2\Delta_{\phi} + 2n$. 
 Factoring out an overall $(x_{13}^2 x_{24}^2)^{-\Delta_{\phi}}$, the conformal block decomposition in the $14 \rightarrow 23$ channel reads
\be
\label{eq:MFTConfBlock}
u^{-\Delta_{\phi}} + 1 + v^{-\Delta_{\phi}} = v^{-\Delta_{\phi}} + v^{-\Delta_{\phi}} \sum_{n, \ell} P_{2\Delta_{\phi} + 2n,\ell} \ \! g_{2\Delta_{\phi} + 2n, \ell} (v, u),
\ee
where the $v^{-\Delta_{\phi}}$ on the RHS is the contribution from the identity operator.  
 If we look at the behavior of the conformal blocks $g_{2\Delta_{\phi} + 2n, \ell} (v, u)$, we notice a simple problem with this equation: it is known that the conformal blocks $g_{2\Delta_{\phi} + 2n, \ell}(v,u)$ in the sum on the RHS each have at most a $\log u$ divergence at small $u$, but the LHS has a $u^{-\Delta_{\phi}}$ divergence.  Thus the LHS cannot be reproduced by any finite number of terms in the sum.  To be a bit more precise, the conformal blocks have a series expansion around $u=0$ with only non-negative integer powers of $u$ and at most a single logarithm appearing, so in particular we can write
\be
v^{-\Delta_{\phi}} g_{2\Delta_{\phi} + 2n, \ell} (v, u) &=& f_0(v) + u f_1(v) + u^2 f_2(v) + \ldots \nn\\
&& + \log(u) \left( \tilde{f}_0 (v) + u \tilde{f}_1 (v) + u^2 \tilde{f}_2 (v) + \ldots \right) .
\ee
But this means that if the sum on the right-hand side of equation (\ref{eq:MFTConfBlock}) converges uniformly, it cannot reproduce the left-hand side, which includes the negative power term $u^{-\Delta_{\phi}}$ and does not include any logarithms.  

The simple resolution of this `paradox' is that the sum over conformal blocks does not converge uniformly near $u = 0$.  In fact, the sum does converge on an open set with positive real $u$, but when Re$[\sqrt u] < 0$ the sum diverges.  So we must define the sum over conformal blocks for general $u$ as the analytic continuation of the sum in the convergent region.  Crucially, the analytic continuation of the sum contains the power-law $u^{-\Delta_{\phi}}$ that is not exhibited by any of the individual terms in the sum.  

Let us see how this works in a bit more detail, so that in particular, we can see that the sum over twists $\tau= 2 \Delta_{\phi} + 2n$ at fixed $\ell$ converges in a neighborhood of $u=0$, but the sum over angular momentum diverges for $u<0$.  For the purpose of understanding convergence, we need only study the conformal blocks when $\tau$ or $\ell$ are very large.  In the very large $\tau$ limit with $|u|, |v| < 1$ the blocks are always suppressed by $u^{\frac{\tau}{2}}$ or $v^{\frac{\tau}{2}}$.  The conformal block coefficients are bounded at large $\tau$ \cite{Pappadopulo:2012jk}.  This means that for small $|u|$ and $|v|$, the sum over $\tau$ will converge.  In fact, once we know that the sum converges for some particular $u_0, v_0$ we see that for $u < u_0$ and $v < v_0$, the convergence at large $\tau$ becomes exponentially faster.  

Now consider the $\ell$ dependence.  At large $\ell$ and fixed $\tau$, we establish in Appendix~\ref{app:LargeLBlocks} that the crossed-channel blocks in the $|u| \ll |v| \ll 1$ limit behave as 
\be
g_{\tau, \ell}(v, u) \approx \frac{\ell^{\frac12} 2^{\tau+2\ell}}{\sqrt{\pi}} v^{\frac{\tau}{2}} K_0\left(2\ell \sqrt{u} \right) \stackrel{\ell \sqrt{u}  \gg 1}{\approx}  2^{\tau + 2\ell - 1} v^{\frac{\tau}{2}} \frac{e^{- 2 \ell \sqrt{u}}  }{\sqrt[4] u} ,
\label{eq:conffactor}
\ee
where $K_0$ is a modified Bessel function. Notice that for Re$[\sqrt u ] > 0$ there is an exponential suppression at very large $\ell$, but for Re$[\sqrt u ] < 0$ there is an exponential growth.  Note also that at small $v$ the lowest twist terms ($n=0$) will dominate. 

Now, the mean field theory conformal block coefficients  in any dimension $d$ are \cite{Unitarity} 
\be
\label{eq:MFTCoeffs}
P_{2\Delta_\phi + 2n, \ell} = \frac{\left[ 1+ (-1)^\ell \right] (\Delta_\phi - \frac{d}{2} + 1)_n^2 (\Delta_\phi)_{n + \ell}^2 }{\ell! n! (\ell+ \frac{d}{2})_n (2 \Delta_\phi + n - d + 1)_n (2 \Delta_\phi + 2n + \ell - 1)_\ell (2 \Delta_\phi + n + \ell - \frac{d}{2})_n},
\ee
where the Pochhammer symbol $(a)_b \equiv \Gamma(a+b)/\Gamma(a)$.  In particular, for $n=0$ and at large even $\ell$ we can approximate
\be
P_{2\Delta_{\phi},\ell} 
\stackrel{\ell \gg 1}{ \approx} q_{\Delta_{\phi}} \frac{\sqrt{\pi} }{2^{2\Delta_{\phi} + 2\ell}} \ell^{2\Delta_{\phi}-\frac32},
\ee
where $q_{\Delta_{\phi}}$ is an $\ell$-independent prefactor.\footnote{Explicitly, $q_{\Delta_{\phi}} =  \left( \frac{8}{ \Gamma(\Delta_{\phi})^2} \right)$. }
Thus the sum in Eq.~(\ref{eq:MFTConfBlock}) at large $\ell$ and $|u| \ll |v| \ll 1$ takes the form
 \begin{equation}
v^{-\Delta_{\phi}} \sum_{n, \,\textrm{large } \ell} P_{2\Delta_{\phi} + 2n,\ell} \ \! g_{2\Delta_{\phi}+2n, \ell} (v, u) \approx  q_{\Delta_{\phi}} 
 \sum_{\textrm{large even} \ \ell}^\infty  \ell^{2\Delta_{\phi}-1} K_0\left(2\ell \sqrt{u}\right).
 \end{equation}
This sum converges at large $\ell$ for positive real $\sqrt u$, and so we will define it by analytic continuation elsewhere in the complex $u$ plane.  As can be easily seen by approximating the sum with an integral, the result reproduces the $u^{- \Delta_{\phi}}$ power-law term on the left-hand side of equation (\ref{eq:MFTConfBlock}), as desired.  Thus we have seen that general power-laws in $u$ are reproduced by conditionally convergent large $\ell$ sums in the conformal block decomposition, with a power-law dependence on $\ell$ producing a related power of $u$ as $u \to 0$.  

\subsection{Existence of Twist $2 \Delta_{\phi} + 2n + \gamma(n, \ell)$ Operators at Large $\ell$}
\label{sec:DblTrace}

In section \ref{sec:MFTExample} we saw how in MFT the sum over large $\ell$ conformal blocks in the crossed $14 \to 23$ channel controls the leading power-law behavior in $u$ in the standard $12 \to 34$ channel.  Now we will use the bootstrap equation (\ref{eq:bootstrap})  to turn this observation into a powerful and general method for learning about the spectrum and the conformal block coefficients $P_{\tau, \ell}$ at large $\ell$ in any CFT.  

Separating out the identity operator, the bootstrap equation reads
\begin{equation}
\label{eq:CrossingBootstrap}
1 +
\sum_{\tau,\ell} P_{\tau, \ell} \ \! u^{\frac{\tau}{2}} f_{\tau,\ell}(u,v) 
=  \p{\frac{u}{v}}^{\Delta_\phi}\p{1+\sum_{\tau,\ell} P_{\tau,\ell} \ \!  v^{\frac{\tau}{2}} f_{\tau,\ell}(v,u)},
\end{equation}
where we have written the conformal blocks as $g_{\tau, \ell}(u,v) = u^{\frac{\tau}{2}} f_{\tau, \ell}(u,v)$ to emphasize their leading behavior at small $u$ and $v$. 
We will work in $d >2$ so that the unitarity bound on twists 
\be
\tau &\ge& 
\left\{
\begin{array}{cc}
\frac{d-2}{2} & (\ell=0),\\
d-2 &  (\ell \ge 1),
\end{array}
\right.
  \label{eq:unitaritybound}
  \ee
strictly separates the identity operator from all other operators.

The arguments in this section will follow from an elementary point: 
\begin{itemize}
\item In the small-$u$ limit, the sum on the right-hand side of equation (\ref{eq:CrossingBootstrap}) must correctly reproduce the identity contribution on the left-hand side.
\end{itemize}
We will show that this implies the existence of towers of operators with increasing spin whose twists approach $2\Delta_\phi+2n$, for each integer $n\geq 0$.  Together with the results in Appendix \ref{app:LargeLBlocks} and the more rigorous arguments in Appendix \ref{app:Rigorous}, we will provide a rigorous proof of this claim.
In subsequent sections, we will consider subleading corrections to the small-$u$ limit of Eq.~(\ref{eq:CrossingBootstrap}) coming from operators of minimal non-zero twist.  

For the remainder of this section we will use the approximate relation
\be
\label{eq:ApproxCrossing}
1 \approx 
 \p{\frac{u}{v}}^{\Delta_\phi}\sum_{\tau,\ell} P_{\tau,\ell} \ \! g_{\tau,\ell}(v,u),
 \qquad
 (u \to 0),
\ee
valid up to strictly sub-leading corrections in the limit $ u \to 0$.\footnote{The sum over conformal blocks on the left-hand side of the crossing relation is necessarily a subleading correction at small $u$.  The reason for this is that we take $v$ to a small but fixed value when we take the small $u$ limit, so the conformal blocks factorize at large spin as shown in Eq.~(\ref{eq:conffactor}) (with $v$ and $u$ interchanged).  The sum over spins on the left-hand side therefore manifestly cannot produce additional singularities in $u$.  The sum over twists is regulated by the $u^{\frac{\tau}{2} - \Delta_\phi}$ factor.} 
As we saw in section \ref{sec:MFTExample}, no finite collection of spins on the right-hand side of (\ref{eq:ApproxCrossing}) can give rise to the left-hand side.  This is true even including an infinite sum over large $\tau$.  To understand how these terms are reproduced, we must study the large $\ell$ region of the sum on the right-hand side of equation (\ref{eq:ApproxCrossing}).  For this purpose we need a formula for the conformal blocks, $g_{\tau, \ell}(v,u)$, at $|u| \ll 1$ and large $\ell$.  

We show in Appendix \ref{app:LargeLBlocks} that the blocks can be approximated in this limit by
\be
\label{eq:fBlock}
g_{\tau, \ell}(v, u) &\approx &
k_{2\ell}(1-z) v^{\tau/2}F^{(d)}(\tau,v)
\qquad(|u|\ll 1\textrm{ and }\ell \gg 1),
\\
k_\beta(x) &\equiv& x^{\beta/2} {}_2F_1(\beta/2,\beta/2,\beta,x),
\ee
where $z$ is defined by $u=z\bar z$, $v=(1-z)(1-\bar z)$, and the function $F^{(d)}(\tau,v)$ is positive and analytic near $v=0$.\footnote{For example, $F^{(2)}(\tau,v)$ is given by Eq.~(\ref{eq:2dFfunction}).  In Appendix \ref{app:LargeLBlocks}, we give recursion relations which allow one to generate $F^{(d)}$ in any even $d$.  In odd $d$, one must resort to solving a differential equation.}  The exact expression for $F^{(d)}$ will not be important for our discussion.  Note that $z\to 0$ at fixed $\bar z$ is equivalent to $u\to 0$ at fixed $v$.

A key feature of Eq.~(\ref{eq:fBlock}) is that the $\ell,z$ dependence of $g_{\tau,\ell}$ factorizes from the $\tau,v$ dependence in the limit $z\to 0,$ $\ell\to\oo$.  Thus, we expect operators with large spin to be crucial for reproducing the correct $z$-dependence on the left-hand side of Eq.~(\ref{eq:ApproxCrossing}).  However, a particular pattern of twists should also be necessary for reproducing the correct $v$-dependence in Eq.~(\ref{eq:ApproxCrossing}).  What is this pattern?

To study this question, it is useful to introduce a conformal block ``density" in twist space,
\be
D_{u,v}(\s) \equiv \p{\frac{u}{v}}^{\Delta_\phi}\sum_{\tau,\ell}P_{\tau,\ell}\de(\tau-\s)g_{\tau,\ell}(v,u).
\ee
By integrating $D_{u,v}(\s)$ against various functions $f(\s)$, we can study the contributions to the crossing equation from operators with different twists.  One should think of it as a tool for studying the spectrum of operators in twist-space.  Since the conformal blocks are positive in the region $0<z,\bar z<1$, and the coefficients $P_{\tau,\ell}$ are positive, $D_{u,v}$ is positive as well.

The full conformal block expansion comes from integrating $D_{u,v}(\s)$ against the constant function $1$.  Thus, the crossing Eq.~(\ref{eq:ApproxCrossing}) in the small $u$ limit reads\footnote{We justify switching the limit and integration in Appendix~\ref{app:Rigorous}.  Roughly, it follows from the fact that the integral of $D_{u,v}(\s)$ over regions with large $\s$ falls exponentially with $\s$.}
\be
\label{eq:crossingEqforrho}
1 &=& \lim_{u\to 0} \int_{d-2}^\oo d\s D_{u,v}(\s) \ \ =\ \ \int_{d-2}^\oo d\s \lim_{u\to 0} D_{u,v}(\s).
\ee

As we discussed above, the $u\to 0$ limit on the RHS is dominated by the sum over large $\ell$, so we are free to substitute the asymptotic form of the blocks Eq.~(\ref{eq:fBlock}) into the definition of $D_{u,v}$ and maintain the same $u\to 0$ (equivalently $z\to 0$) limit,
\be
\lim_{u\to 0} D_{u,v}(\s) &=& \p{\lim_{z\to 0} z^{\Delta_{\phi}}\sum_{\tau,\ell}P_{\tau,\ell} k_{2\ell}(1-z)  \de(\tau-\s)} v^{\frac{\s}{2}-\Delta_{\phi}}(1-v)^{\De_\phi}F^{(d)}(\s,v),
\ee
where we have used $\bar z = 1-v+O(z)$.
Note that after this substitution, the $\tau$ and $v$-dependence factors out into an overall function $v^{\frac{\s}{2}-\Delta_{\phi}}(1-v)^{\De_\phi}F^{(d)}(\s,v)$, while the $u$ and $\ell$ dependence is encapsulated in a particular weighted sum of OPE coefficients,
\be
\label{eq:definitionofrhobody}
\rho(\s) &\equiv& \lim_{z \to 0}z^{\Delta_{\phi}}\sum_{\tau,\ell}P_{\tau,\ell}k_{2\ell}(1-z)  \de(\tau-\s) .
\ee

By crossing symmetry Eq.~(\ref{eq:crossingEqforrho}), the density $\rho(\s)$ satisfies
\be
\label{eq:integralcrossingrelation}
1&=&(1-v)^{\Delta_{\phi} }\int_{d-2}^\oo d\s \rho(\s)  v^{\frac{\s}{2}-\Delta_{\phi}}F^{(d)}(\s,v) .
\ee
We claim that the only way to solve this relation with positive $\rho(\s)$ is if $\rho$ is given by its value in mean field theory, namely a sum of delta functions at even integer-spaced twist:
\be
\label{eq:MFTclaim}
\rho(\s) &=& \sum_{n=0,1,\dots} P^\mathrm{MFT}_{2\Delta_{\phi}+2n}\de(\s-(2\Delta_{\phi}+2n)).
\ee
where the mean field theory coefficients were given as a function of $n$ and $\ell$ in equation (\ref{eq:MFTCoeffs}).  Expanding these coefficients at large $\ell$ and performing the sum in Eq.~(\ref{eq:definitionofrhobody}) gives\footnote{To perform this computation explicitly, we use the fact that the sum is dominated by the region of fixed $z\ell^2$, so that one may use the approximation of Eq.~(\ref{eq:FixedUApprox}).}
\be
\label{eq:MFTAccumulated}
P^{MFT}_{2\Delta_\phi + 2n} = \frac{1}{2^{2\Delta_{\phi}+2n}}
\frac{ (\Delta_\phi- \frac{d}{2}+1)_n^2 }{  n!  (2 \Delta_\phi + n - d + 1)_n }.
\ee

Let us give a brief argument for why this is the case.  Note that the function $F^{(d)}(\s,v)$ is analytic and positive near $v=0$, so the small $v$ behavior of the above integral comes from the term $v^{\frac{\s}{2}-\Delta_{\phi}}$.  Since the right-hand side is independent of $v$, and the density $\rho(\s)$ is nonnegative, we see that $\rho(\s)$ must be zero for $\s<2\Delta_{\phi}$ and also have a contribution proportional to $\de(\s-2\Delta_{\phi})$.  

The necessity of the other terms $\de(\s-(2\Delta_{\phi}+2n))$ now follows from the fact that the LHS is independent of $v$, while the first term $\de(\s-2\Delta_{\phi})$ contributes a power series in $v$ about $v=0$, namely $F^{(d)}(2\Delta_{\phi},v)$.  Terms with $n> 0$ are needed to cancel successive powers of $v$ from this first term.  To see the necessity of this result most clearly, it is helpful to subtract the contribution of $\de(\s-2\Delta_{\phi})$ from both sides of Eq.~(\ref{eq:integralcrossingrelation}):
\be
O(v) &=& (1-v)^{\Delta_{\phi} }\int_{d-2}^\oo d\s \rho'(\s)  v^{\frac{\s}{2}-\Delta_{\phi}}F^{(d)}(\s,v)
\ee
where $\rho'(\s)=\rho(\s)-P^\mathrm{MFT}_{2\Delta_{\phi}}\de(\s-2\Delta_{\phi})$. We are left with an $O(v)$ term on the LHS which must now be matched by a $\de(\s-(2\Delta_{\phi}+2))$ term in $\rho'$.  Repeating this algorithm iteratively, one can fix $\rho(\s)$ to be given by its value in MFT.\footnote{We also note that one might reproduce this conclusion directly by doing a projection of the LHS and RHS onto specific high-spin conformal blocks using the method of ``conglomerating'' \cite{Unitarity}.}

The result Eq.~(\ref{eq:MFTclaim}) has several consequences.  Firstly, it implies the existence of 
a tower of operators with increasing spin whose twists approach
$\tau=2\Delta_{\phi}+2n$, for each $n\geq 0$.  To see this, let us integrate Eq.~(\ref{eq:MFTclaim}) over a bump function $h_\epsilon(\s)$ with some width $\epsilon$ around $\s=2\Delta_{\phi}+2n$.  Using the definition in Eq.~(\ref{eq:definitionofrhobody}), we obtain\footnote{We justify interchanging the limit and integration in Appendix \ref{app:Rigorous}.}
\be
\lim_{z\to 0} z^{\Delta_{\phi}}\sum_{\tau,\ell}P_{\tau,\ell}h_\epsilon(\tau)  k_{2\ell}(1-z)  &=& P_{2\Delta_{\phi}+2n}^\mathrm{MFT}.
\ee
The limit vanishes termwise on the LHS, so a finite result can only come from the sum over an infinite number of terms.  Thus, for any $\epsilon$, there are an infinite number of operators with twist $\tau=2\Delta_{\phi}+2n+O(\epsilon)$.

We can also be more precise about the contribution of these operators to the conformal block expansion at large $\ell$.  The sum above can be written as an integral over the OPE coefficient density in $\ell$, for operators with twist near $2\Delta_{\phi}+2n$
\be
\sum_{\tau\sim 2\Delta_{\phi}+2n,\ell} P_{\tau,\ell} k_{2\ell}(1-z) &=& \int_0^\oo d\ell f_n(\ell) k_{2\ell}(1-z),
\label{eq:Fdef}
\ee
where we define an OPE coefficient density
\be
\label{eq:fLDefinition}
f_n(\ell) \equiv
\sum_{\tau\sim 2\Delta_{\phi}+2n,\ell'} P_{\tau,\ell'} \de(\ell-\ell').
\ee
For simplicity, we no longer show the bump function $h_\epsilon(\tau)$ explicitly, but indicate its presence by writing $\tau\sim 2\Delta_{\phi}+2n$.  

One intuitively expects that the OPE coefficient density $f_n(\ell)$ must be constrained at large $\ell$ in order to reproduce the identity operator in the crossed channel.  In mean field theory $f_n(\ell)$ has a power-law behavior, and so we expect that, in an averaged sense, $f_n(\ell)$ must be similar in any CFT.  This motivates introducing an integrated density
\be
F_n(L) \equiv \int_0^L d \ell \frac{\Gamma(2 \ell)}{\Gamma(\ell)^2}  f_n(\ell) .
\ee
In Appendix \ref{app:BoundsF} we prove both upper and lower bounds
\be
A_U L^{ 2 \Delta_{\phi} }  \ > \ F_n(L)  \ > \ A_L \frac{ L^{2 \Delta_{\phi} } }{\log(L)}
\ee
for some coefficients $A_U$ and $A_L$ in the limit of very large $L$.  We expect that the lower bound can be improved to eliminate the logarithm and make a prediction $F_n(L) =A_n L^{2\Delta_\phi}$.\footnote{ In the case where the functions $k_\beta(1-z)$ are replaced by their exponential approximation Eq.~(\ref{eq:Exponentialdecay}), the Hardy-Littlewood Tauberian theorem says that the upper and lower bound at large $L$ are the same, and it fixes their coefficient. It seems likely to us that an analogous theorem could be proven for the case at hand. } We calculate in Appendix \ref{app:BoundsF} that such a prediction would necessarily fix $A_n$ to be
\be
\lim_{L \rightarrow \infty} L^{-2\Delta_\phi} F_n(L) = \frac{P^{MFT}_{2\Delta_\phi + 2n} }{\Delta_\phi \Gamma(\Delta_\phi)^2 }.
\ee

In summary, we have shown that in any CFT we must have operators accumulating at twists $2 \Delta_{\phi} + 2n$ at large $\ell$.  In simple cases where these accumulation points are populated by a single operator at each $\ell$, as in all perturbative theories, we can obtain specific relations for sums over the anomalous dimensions $\gamma(n, \ell)$ and the conformal block coefficients $\delta P_{2 \Delta_{\phi}+2n, \ell}$.  We will explore these relations in the next subsection. 
In Appendix \ref{app:DistinctOperators} we briefly explain how these results generalize when we have distinct operators $\phi_1$ and $\phi_2$, so that there must exist operators accumulating at twist $\Delta_1 + \Delta_2 + 2n$ as $\ell \to \infty$.

\subsubsection{Relation to Numerical Results and the 3d Ising Model}

In \cite{ElShowk:2012ht}, the authors used numerical boostrap methods to constrain the dimensions of operators appearing in the OPE of a scalar $\phi$ with itself in 3d CFTs.  They found numerical evidence that the minimum twist $\tau_\ell$  at each spin $\ell$ in the $\phi\x\phi$ OPE satisfies
\be
\label{eq:lowesttwistbound}
\tau_{\ell} &\leq& 2\Delta_\phi.
\ee
(More precisely, they compute a series of numerical bounds on $\tau_{\ell}$, which appear to approach the presumably optimal bound Eq.~(\ref{eq:lowesttwistbound}), at least for $\ell=2,4,6$.)  Here, we note that Eq.~(\ref{eq:lowesttwistbound}) follows from our results, together with Nachtmann's theorem \cite{Nachtmann:1973mr}.  We have seen, among other things, that there exist operators in the $\phi\x\phi$ OPE with twist arbitrarily close to $2\Delta_\phi$ and arbitrarily high spin $\ell$.  Nachtmann's theorem states that $\tau_{\ell}$ is an increasing function of $\ell$ for $\ell>0$, which implies Eq.~(\ref{eq:lowesttwistbound}) for all $\ell>0$.

The 3d Ising Model is a particularly interesting example in light of this result.  The theory contains a scalar $\s$ with dimension $\Delta_\s\approx 0.518$.  Thus, the minimum twist operators at each $\ell$ have twist less than $\approx 1.036$, which is very close to the unitarity bound.  One might interpret these operators for $\ell\geq 4$ as approximately conserved higher-spin currents.  It would be interesting to understand what the existence of these approximate currents implies for structure of the theory.

\subsection{Properties of Isolated Towers of Operators}
\label{sec:IsolatedTowers}

So far, we have made no assumptions about the precise spectrum of operators that accumulate at the special values $\tau\sim 2\Delta_{\phi}+2n$.  However, it is interesting to consider the case where an accumulation point $2\Delta_{\phi}+2n$ is approached by a single tower of operators $\mathcal{O}_{2 \Delta_\phi + 2n,\ell}$ with $\ell=0,2,\dots$, which are separated by a twist gap from other operators in the spectrum at sufficiently large spin.  This occurs in virtually every example we are aware of, including all theories with a perturbative expansion parameter such as $1/N$, and we will also discuss some non-perturbative examples at the end of this subsection.  It would be very interesting to identify any CFTs without operators with twists near $2 \Delta_\phi + 2n$ for every sufficiently large value of $\ell$.

With this additional assumption, we will be able to characterize subleading corrections to the bootstrap equation, Eq.~(\ref{eq:bootstrap}), in the small $u$ limit.  On the left-hand side, these corrections come from the operators $\CO_m$ with minimal nonzero twist.
Thus, we have the approximate relation
\be
\label{eq:BetterApproxCrossing}
1 + \sum_{\ell_m=0}^2 P_{m} u^{\frac{\tau_{m}}{2}} f_{\tau_{m},\ell_{m}}(0,v) \approx 
 \sum_{\tau,\ell} P_{\tau,\ell} \ \!  v^{\frac{\tau}{2}- \Delta_{\phi}} u^{\Delta_{\phi}} f_{\tau,\ell}(v,u),
\ee
which is valid up to subleading corrections in $u$ in the limit $ u \to 0$.  We have assumed that $\ell_m \leq 2$ because higher spin operators either have twist greater than that of the energy-momentum tensor or, as argued in \cite{Maldacena:2011jn, Maldacena:2012sf}, they are part of an infinite number of higher-spin currents  that couple as if they were formed from free fields.\footnote{Strictly speaking, the arguments in \cite{Maldacena:2011jn, Maldacena:2012sf} assumed $d=3$ and studied correlators of currents, but it is likely that they can be extended to $d\ge 3$.  In any case, one can view $\ell_m \leq 2$ as an assumption. }  
Dolan and Osborn \cite{Dolan:2011dv} have given a formula in general $d$ appropriate for the conformal blocks corresponding to $\CO_m$ exchange on the left-hand side, where we have expanded at small $u$.  This is
\be
\label{eq:TauMinBlocks}
f_{\tau_{m},\ell_{m}}(0,v) =  (1-v)^{\ell_m} {}_2F_1\left( \frac{\tau_m}{2} +\ell_m, \frac{\tau_m}{2} + \ell_m, \tau_m+2 \ell_m, 1-v\right).
\ee
It will be important that this hypergeometric function can be expanded in a power series at small $v$ with terms of the form $v^k (a_k + b_k \log v)$.  The logarithms will be related to the `anomalous dimensions' that emerge at large $\ell$.

Now we expect there to be a finite separation between the lowest twist $\tau_m$ and the other twists in the theory.  To prove this, we consider two cases separately, that of $\tau_m < d-2$ and that of $\tau_m \ge d-2$. 
In the former case, $\CO_m$ must be a scalar operator due to the unitarity bound (\ref{eq:unitaritybound}). We will assume that there are a finite number of scalar operators with dimension below any given value,\footnote{This assumption would follow, for instance, from the assumption that the CFT has a well-defined partition function at non-zero temperature, or that the four-point function of the energy-momentum tensor is finite.} which immediately implies that the twist of $\CO_m$ must have a finite separation from the other twists in the theory.  To have a non-vanishing 4-pt correlator $\phi$ must be uncharged, so in the absence of lower twist scalars we have $\tau_m=d-2$, because the energy-momentum tensor will appear in the $\phi(x) \phi(0)$ OPE.   We can then apply the Nachtmann theorem, which says that minimal twists must be non-decreasing functions of $\ell$, to conclude that $\tau_m$ is separated from the other twists in the theory.  Note that, crucially, unless $\phi$ is a free scalar field we have $\frac{\tau_m}{2} - \Delta_{\phi} < 0$, so the sub-leading powers of $u$ grow as $u \to 0$.  Taken together, these comments imply that there is a limit $u \to 0$ where the exchange of the identity operator plus a finite number of $\CO_m$ dominates the left-hand side of equation (\ref{eq:CrossingBootstrap}).

Assuming the existence of an operator with twist approaching $2 \Delta_\phi + 2n$ for each $\ell$, we would like to constrain the deviation of their conformal block coefficients $\delta P_{2 \Delta_{\phi}+2n, \ell}$ from MFT and their anomalous dimensions  $\g(n,\ell) \equiv \De_{\mathcal{O}_{n,\ell}}-2\Delta_{\phi}-2n-\ell$.  This is possible because the $\CO_m$ contribute a dominant sub-leading contribution at small $u$, with a known $v$-dependence that can be expanded in a power series with integer powers at small $v$.  The fact that we have only integer powers $v^n$ and $v^n \log v$ multiplying  $u^{\frac{\tau_m}{2}}$ on the left-hand side of equation (\ref{eq:BetterApproxCrossing}) means that  the right-hand side can reproduce these terms only with the conformal blocks we just discovered above, namely those with twists approaching the accumulation points $\tau(n, \ell) = 2 \Delta_{\phi} + 2n + \gamma(n, \ell)$ so that $v^{\frac{\tau(n,\ell)}{2} - \Delta_{\phi}}$ is an integer power in the $\ell \to \infty$ limit.  

In fact, expanding Eq.~(\ref{eq:TauMinBlocks}) for the $\CO_m$ conformal blocks at small $v$ gives
\begin{equation}
f_{\tau_m, \ell_m}(v) = \frac{\Gamma(\tau_m + 2 \ell_m)(1-v)^{\ell_m} }{\Gamma^2 \left(\frac{\tau_m}{2} + \ell_m \right)}  
\sum_{n=0}^\infty \left( \frac{\left(\frac{\tau_m}{2} + \ell_m \right)_n}{n!} \right)^2 v^n \left[ 2 \left(\psi(n+1) - \psi \left( \frac{\tau_m}{2} + \ell_m \right) \right)- \log v \right] ,
\label{eq:ExplicitTauMinBlock}
\end{equation}
where $\psi(x)=\Gamma'(x)/\Gamma(x)$ is the Digamma function, and $(a)_b = \Gamma(a+b)/\Gamma(a)$ is the Pochhammer symbol.  The details of this formula are not especially important, except insofar as it makes explicit the connection between the coefficients of $v^n$ and $v^n \log v$ in the series expansion.  We will now see that the $v^n$ terms must come from $\delta P_{2 \Delta_{\phi}+2n, \ell}$ while the $v^n \log v$ terms are a consequence of $\gamma(n, \ell)$.  This means that the `anomalous dimension' and the correction to the conformal block coefficients must be related at large $\ell$.  These quantities were seen  to be related \cite{JP} to all orders in perturbation theory \cite{Unitarity} in the presence of a $1/N$ expansion, so our result extends this relation to a non-perturbative context.

The $v^n \log v$ terms in equation (\ref{eq:ExplicitTauMinBlock}) can only be reproduced by expanding $v^{\frac{\tau}{2} - \Delta_{\phi}}$ in $\gamma(n, \ell)$ in the large $\ell$ conformal blocks.  For simplicity let us consider the situation where there is only one operator accumulating near $2 \Delta_{\phi} + 2n$ for each $\ell$, and that the conformal block coefficients approach $P^{MFT}_{2 \Delta_{\phi}+2n, \ell}$.  In this case we can write the RHS of the crossing relation as 
\be
\sum_{\ell} P^{MFT}_{2 \Delta_{\phi}+2n, \ell} \left[ \frac{\gamma(n, \ell)}{2} \log v \right]  \frac{\ell^{\frac12} 4^{\ell}}{\sqrt{\pi}} z^{\Delta_{\phi}} K_0(2\ell \sqrt z)   v^{n} (1-v)^{\Delta_{\phi}} F^{(d)}(2 \Delta_{\phi} + 2n, v) ,
\ee
where we have used a Bessel function approximation to $k_{2\ell}(1-z)$ discussed in Appendix \ref{app:Bessel}.\footnote{Strictly speaking, the Bessel function approximation breaks down for $\ell \gg 1/\sqrt{z}$, so we are here implicitly using the fact that the sum including the MFT coefficients is dominated by the region of fixed $\ell^2 z$, where the approximation is valid. If the reader is concerned about this, one can instead write these sums using the hypergeometric function $k_{2\ell}(1-z)$. However, we find the Bessel function formulae to be useful for explicitly doing computations using the integral approximations to these sums.}
In order for the sum to produce an overall factor of $u^{\frac{\tau_m}{2}} \sim z^{\frac{\tau_m}{2}}$, we must have power law behavior in $\gamma(n,\ell)$ at very large $\ell$,
\be
\gamma(n, \ell) = \frac{\gamma_n}{\ell^{\tau_m}},
\ee
with a coefficient $\gamma_n$ related to the OPE coefficient $P_m$ of the leading twist operator $\CO_m$ in equation (\ref{eq:BetterApproxCrossing}). In large $N$ theories, the coefficients $\gamma_n$ are suppressed by powers of $1/N$ as discussed in \cite{Alday:2007mf}. However, we stress that all we need in order to expand $v^{\frac{\gamma(n,\ell)}{2}}$ in $\log v$ in the large $\ell$ sum is the property that it is power law suppressed as $\ell \rightarrow \infty$, which is true even if the coefficients $\gamma_n$ are $O(1)$ or larger.

The integer powers of $v$ in equation (\ref{eq:ExplicitTauMinBlock}) must then be reproduced by 
\be
\sum_{\ell} P^{MFT}_{2 \Delta_{\phi}+2n, \ell} \left[ \delta P_{2 \Delta_{\phi}+2n, \ell}  + \frac12 \gamma(n,\ell) \frac{d}{dn} \right] v^n \frac{\ell^{\frac12} 4^{\ell}}{\sqrt{\pi}} z^{\Delta_{\phi}} K_0(2\ell\sqrt z)  (1-v)^{\Delta_{\phi}} F^{(d)}(2 \Delta_{\phi} + 2n, v) ,
\ee
where the $\gamma(n,\ell) \frac{d}{dn}$ piece comes from expanding the $\tau$ dependence of the conformal block in small $\gamma(n,\ell)$. Again requiring that we correctly produce the overall $u^{\frac{\tau_m}{2}} \sim z^{\frac{\tau_m}{2}}$ behavior, we must have
\be
\delta P_{2 \Delta_{\phi}+2n, \ell} = \frac{c_n}{\ell^{\tau_m}}
\ee
to leading order in $1/\ell$ in the large $\ell$ limit, with a coefficient $c_n$ related to $\gamma_n$. 

As an example, it is particularly simple to do this matching explicitly for the leading twist tower with $n=0$. In this case, matching the $\log v$ and $v^0$ terms gives the relations 
\be\label{eq:n0matching}
\gamma_0 = -P_{m} \frac{2\Gamma(\Delta_{\phi})^2 \Gamma(\tau_m+2 \ell_m)}{ \Gamma(\Delta_{\phi}-\frac{\tau_m}{2})^2\Gamma(\frac{\tau_m}{2}+\ell_m)^2} \, , \,\,\, c_0 = \left[\psi \left(\frac{\tau_m}{2}+\ell_m \right) +  \gamma +\log 2  \right] \gamma_{0}, 
\ee 
where $\gamma$ is the Euler-Mascheroni constant. It is important to note the relative sign between $\gamma_0$ and $P_m$ (which is strictly positive), since this is required in order to satisfy the Nachtmann theorem asymptotically at large $\ell$. It is then straightforward to continue this matching to higher orders in $v$.

\subsubsection{Implications for SCFTs in 4d}

In \cite{Poland:2010wg, Poland:2011ey} the crossing relation was examined for a chiral primary operator $\Phi$ of dimension $\Delta_\phi$ in an $\mathcal{N}=1$ SCFT, and it was observed that in the OPE of $\Phi(x) \Phi(0)$ there exists an infinite tower of operators with spin $\ell$ and dimension exactly $2 \Delta_\Phi + \ell$.  Supersymmetry and unitarity protect the dimensions of these operators, and further require a gap in twist before non-protected operators can appear.  Thus, these operators form an isolated tower with vanishing anomalous dimensions.  This immediately implies that the correlator in the $\Phi^\dag \Phi \to \Phi \Phi^{\dag}$ channel must satisfy 
\be
\langle \Phi^\dag(x_1) \Phi(x_2)  \Phi(x_3) \Phi^{\dag}(x_4) \rangle = \frac{1}{(x_{12}^2 x_{34}^2)^{\Delta_{\phi}}} 
\left( 1+ u^{1} \left(0 \log v + c  + \ldots \right) + \ldots \right)
\ee
where the $\ldots$ denote higher order terms in $u$ and $v$, and the power $u^{1}$ comes from the $U(1)_R$ current multiplet (containing the stress tensor), which has $\frac{\tau}{2} = 1$. The point is that the term $u^{1} \log v$ must be absent, because it could only arise from the anomalous dimensions of operators with twist $2\Delta_\phi + \gamma(\ell)$ in the $\Phi \Phi \to \Phi^\dag \Phi^\dag$ channel, but we know that due to supersymmetry, $\gamma(\ell) = 0$ exactly.  The explicit results of \cite{Poland:2010wg} show that the superconformal block relevant for the $\Phi^\dag \Phi \to \Phi \Phi^{\dag}$ channel is\footnote{Our normalization for the blocks removes a factor of $(-\frac{1}2)^{\ell}$ compared to that used in \cite{Poland:2010wg}.}
\be
\mathcal{G}_{\tau=2, \ell_m}(u, v) = (-1)^{\ell_m}\left[g_{2, \ell_m}(u,v) - \left( \frac{\ell_m + 1}{4 \ell_m + 6}  \right) g_{2, \ell_m+1}(u,v) \right] 
\ee
in terms of the usual 4d blocks given in equation (\ref{eq:Standard4dBlocks}).  Taking $\ell_m = 1$ for the $U(1)_R$ current, one can easily verify that the $u \log v$ term cancels in this linear combination of conformal blocks in the limit of small $u$ and $v$.\footnote{Note that if instead one considers the s-channel expansion of the correlator $\langle \Phi^\dag \Phi  \Phi^{\dag} \Phi \rangle$ then there is no longer a relative sign between the even and odd spins in the superconformal block \cite{Poland:2010wg}, so a $u^{1}\log v$ term is present.  However, the conformal block expansion in the s-channel cannot be immediately compared to the $\Phi \Phi \rightarrow \Phi^\dagger \Phi^\dagger$ channel because passing between these two different OPE limits requires changing the radial ordering of operators, which introduces phases $\sim (-1)^\ell$ from crossing branch cuts. 
}  In the case where there are non-$R$ currents in the $\phi \times \phi^\dagger$ OPE, these currents would also appear in multiplets that contain scalar components with $\frac{\tau}{2} = 1$.  Consequently, the cancellation also has to occur for $\ell_m=0$, as one can easily verify in the blocks themselves.\footnote{More generally, there exist theories with an infinite number of higher-spin currents, and the anomalous dimensions of the $n=0$ operators should be protected in such cases as well. Additionally, while twists greater than 2 would not be the minimal twist and therefore not obviously constrained by our results, the presence of $u^{\frac{\tau}{2}} v^0 \log v$ terms would at the very least impose non-trivial constraints that would have to be satisfied to be consistent with vanishing anomalous dimensions in the cross-channel. In any case, any fears of a possible contradiction are readily allayed: it is easy to verify that in fact the $u^{\frac{\tau_m}{2}} v^0\log v$ terms in the ${\cal G}_{\tau_m, \ell_m}$ super-conformal blocks cancel for any $\ell_m$ and any $\tau_m$. } This provides a non-trivial consistency check of our results and those of  \cite{Poland:2010wg}.

We can proceed to consider the OPE coefficients of the twist $2 \Delta_\phi$ tower, which were bounded as a function of $\Delta_\phi$ in \cite{Poland:2011ey}.  Our results predict that these should approach the mean field theory conformal block coefficients at a rate $\ell^{-2}$, and this rate of convergence could easily be matched to bounds from the numerical bootstrap in the future.

\subsection{Bounding Contributions from Operators with General Twists}
\label{sec:Bound}

Finally, let us show that as $\ell \to \infty$, the contribution from accumulation points $\tau_a$ other than $2 \Delta_{\phi} + 2n$ is strictly bounded.  An analogous generalization for distinct operators follows from observations in Appendix \ref{app:DistinctOperators}.  The idea of the argument is very simple -- specific power-law behaviors in $\ell$ for the conformal block coefficients $P_{\tau, \ell}$ result in related power-law contributions at small $u$.  Since we explicitly know the leading and sub-leading behavior as $u \to 0$, we can obtain a bound on the conformal block coefficients using the crossing symmetry relation, Eq.~(\ref{eq:BetterApproxCrossing}).  The remainder of this subsection formalizes these claims.

Consider all terms on the right-hand side of (\ref{eq:BetterApproxCrossing}) at large $\ell$ with $|\tau - (2 \Delta_{\phi} + 2n)| > \epsilon$ for some $\epsilon > 0$ small but fixed.  This bound separates out the contributions we studied in the previous subsections. Furthermore, let us consider only operators with twists $\tau < \tau_*$ for some arbitrary choice of $\tau_*$.  The reason for imposing this bound on $\tau$ is that we wish to constrain the CFT spectrum and the conformal block coefficients at large $\ell$, and by this we mean large $\ell$ with fixed $\tau$.  In the analogy with scattering, we are studying the scattering amplitude at large impact parameter and fixed center of mass energy. 

Let us define a quantity that is the partial sum of the right-hand side of (\ref{eq:BetterApproxCrossing}) keeping only operators with $\tau<\tau_*, \ell > \ell_*\gg 1/\sqrt{z}$, and $|\tau - (2\Delta_{\phi}+2n)| > \epsilon$:
\be
\textrm{RHS}(\tau_*, \ell_*) &\equiv&  \sum_{\substack{\tau< \tau_*,\ell> \ell_* \\ \tau \ne 2\Delta_{\phi} + 2n \pm O(\epsilon)}} P_{\tau,\ell} \ \!  v^{\frac{\tau}{2}- \Delta_{\phi}} u^{\Delta_{\phi}} f_{\tau,\ell}(v,u).
\ee 
Then we can approximate 
\be
\textrm{RHS}(\tau_*, \ell_*) \stackrel{\ell_* \gg 1/\sqrt{z}}{\approx}  z^{\Delta_{\phi}} \sum_{\tau < \tau_*, \ell>\ell_*} P_{\tau, \ell} k_{2\ell}(1-z)  \left[ v^{\frac{\tau}{2} - \Delta_{\phi}}  (1-v)^{\Delta_{\phi}} F^{(d)}(\tau, v) \right]   .
\ee
The idea will be to combine together all the various values of $\tau$ for each $\ell$.  Since the conformal block coefficients satisfy $P_{\tau, \ell} > 0$ by unitarity, a weighted sum of them will also be positive.  Furthermore, if we can bound their weighted sum then we can bound each individual term.  For all physical $\tau \leq \tau_*$ the function $v^{\frac{\tau}{2}-\Delta_{\phi}}  (1-v)^{\Delta_{\phi}} F(\tau, d, v)$ will be bounded from above by some $B(\tau_*, d, v)$, so we can write an inequality
\be
\textrm{RHS}(\tau_*, \ell_*)< B(\tau_*, d, v)   z^{\Delta_{\phi}} \sum_{\tau < \tau_*, \ell>\ell_*} P_{\tau, \ell} k_{2\ell}(1-z) .
\ee
For each value of $\ell$, there can be only a finite number of operators with $\tau < \tau_*$.  This means that we can define a new quantity that includes the contributions of all these operators at fixed $\ell$:
\be
Q_{\tau_*, \ell} \equiv  \sum_{\tau < \tau_*} P_{\tau, \ell}  .
\ee
Now we have the bound
\be
\textrm{RHS}(\tau_*, \ell_*) < B(\tau_*, d, v) z^{\Delta_{\phi}} \sum_{\ell>\ell_*} Q_{\tau_*, \ell} k_{2\ell}(1-z).
\ee
Again, the lower-bound $\ell_*$ can be taken arbitrarily large since only the infinite sum over $\ell$ produces additional $u^{-1}$ singularities; operators that do not belong to an infinite tower of spins are irrelevant. 

Now, for the purposes of this argument we can also approximate the $\ell > \ell_*$  sum by an integral.  In order to avoid producing non-integer powers of $v$ on the LHS of (\ref{eq:BetterApproxCrossing}), we must then have that
\be
\lim_{z \to 0} \left[ z^{\Delta_{\phi} - \frac{\tau_{m}}{2} } \int^\infty_{\ell_*} d \ell \ \! Q_{\tau_*, \ell} k_{2\ell}(1-z) \right] 
= 0 .
\ee
If we use the $K_0$ approximation of Eq.~(\ref{eq:FixedUApprox}), then performing a change of variables to $y = \ell \sqrt{z}$ immediately shows that since $Q_{\tau_*, \ell} > 0$, we expect to have an asymptotic bound $Q_{\tau_*, \ell} < 4^{-\ell} \ell^{2 \Delta_{\phi} -\frac32 - \tau_m} $ in the large $\ell$ limit, at least in an averaged sense when we smear over a large number of $\ell$.  More precisely, we can use the arguments in Appendix \ref{app:BoundsF} to show that 
\be
\int^L_{\ell_*} d \ell  \frac{\Gamma(2\ell)}{\Gamma(\ell)^2} Q_{\tau_*, \ell} < A \ \! L^{2 \Delta_\phi - \tau_m }
\ee
for some positive constant $A$ at very large $L$.  This provides a general smeared bound for every sequence of $P_{\tau, \ell}$ as $\ell \to \infty$.  Note, however, that our method cannot strictly exclude examples where large but extremely rare conformal block coefficients occasionally appear at large $\ell$.

\subsection{Failure in Two Dimensions}
\label{sec:2d}

In the previous sections, we had to restrict to $d>2$ dimensions in order to have a gap between the twist of the identity operator and $\tau_m$. 
It is illuminating to see how the absence of such a gap in $d=2$ theories explicitly leads to violations of our conclusions in specific examples.  We will focus here on the simplest of such examples, the $c=\frac12$ minimal (i.e. $d=2$ Ising) model (see e.g. \cite{Ginsparg} for a review).  This theory contains three Virasoro primary operators, all scalars:  $1,\sigma,$ and $\epsilon$, of dimensions $0,\frac{1}{8}$, and $1$ respectively, as well as all their Virasoro descendants.  Consider the operator $\sigma$; the $\sigma \times \sigma$ OPE in this case can be summarized succinctly as
\be
[\sigma] [\sigma] &=& [1] + [\epsilon],
\ee
where $[\CO]$ denotes the full Virasoro conformal block associated with an operator.  Now, since $\epsilon$ and 1 both have integer dimensions, and the Virasoro operators just raise the dimension by integers, this means that every  operator that appears in the $\sigma$ conformal block decomposition has integer twist, violating our conclusion in $d>2$ that there must be operators with twist $\tau = 2\Delta_\sigma +2n = \frac{1}{4} + 2n$.  To see what has gone wrong, examine the bootstrap equation in this theory at $|u|\ll |v| \ll 1$:
\be
u^{-\frac{1}{8}} + \sum_{ \ell} P_{0, \ell} u^{-\frac{1}{8}} f_{0, \ell}(u,v)  =
 \sum_{\tau, \ell} P_{\tau, \ell} v^{\frac{\tau}{2} - \frac{1}{8}} f_{\tau, \ell}(v, u) + \textrm{subleading in } 1/u .
\ee 
In this case there is no gap between the twist of the identity operator and $\tau_m$.  Furthermore, our assumption from the analysis of \cite{Maldacena:2011jn, Maldacena:2012sf} that there is no non-trivial infinite tower of conserved higher-spin currents with $\tau(\ell) = d-2$ for $\ell > 2$ is also violated.  Far from having an isolated dominant contribution from the identity operator at small $u$ followed by a finite number of isolated contributions from twist $\tau_m$ operators (followed by everything else), we immediately have an infinite tower of contributions all at $\tau=0$.  Now we see why there are no operators in this theory with twist $\tau=2\Delta_\sigma$: the existence of this low-twist tower means that the identity operator can be (and is) cancelled by contributions {\em on the same side of the crossing relation}. In fact, in this case, the $\tau=0$ tower contributes not only the same $u^{-\Delta_\sigma}$ singularity, but it also contributes a $v^{-\Delta_\sigma}$ coefficient, for a total of $(u v)^{-\Delta_\sigma}$.  The resulting singularity in the cross-channel can be seen explicitly in the exact four-point function, which contains a leading singularity at small $u$ and $v$ of the form
\be
G_\sigma(z,\bar{z}) &\sim& \frac{1}{(u v)^{\Delta_\sigma}},
\ee
as opposed to the usual $u^{-\Delta_\sigma}$.  It is interesting to note that the constraints from the Virasoro algebra that make many $d=2$ CFTs solvable also directly cause them to differ quite drastically in their behavior at large spin from essentially all other CFTs.

\section{AdS Interpretation}
\label{sec:super}

To the uninitiated, results concerning the CFT spectrum and conformal block coefficients may appear rather technical.  However as recent work has shown \cite{Katz, ScatteringStates, Analyticity}, both anomalous dimensions and OPE or conformal block coefficients have a very simple interpretation as amplitudes for scattering processes in AdS space.  This follows from the fact that in global AdS, time translations are generated by the dilatation operator $D$ of the dual CFT, so anomalous dimensions in the CFT represent energy shifts of bulk states due to interactions.  By the Born approximation, these are related to scattering amplitudes in the perturbative regime \cite{Katz}.  A thorough investigation of this connection in the context of gravitational scattering in AdS at large impact parameter was performed in \cite{ourEikonal, ourCPW}, and in Appendix \ref{app:Joao} we explicitly compare our results to theirs in the region of overlap.

To understand the connection to AdS, consider any scalar primary operator $\phi$ with dimension $\Delta_{\phi}$, which creates a state $|\phi \rangle = \phi | 0 \rangle$ when acting on the vacuum of the CFT.  If we were working at large $N$ and $\phi$ was single-trace, then we could interpret $|\phi \rangle$ as a single-particle state in AdS.  Furthermore, we could interpret the operators $\CO_{\tau, \ell}$ appearing in the OPE 
\be
\phi(x) \phi(0) = \sum_{\tau, \ell} c_{\tau,\ell} f_{\tau,\ell}(x,\partial) \CO_{\tau, \ell}(0)
\ee
of $\phi$ with itself as 2-particle states whose anomalous dimensions were due to bulk interactions.  The operators $\CO_{\tau, \ell}$ at large $\ell$ correspond to states with large angular momentum in AdS, so that the two particles are orbiting a common center with a large angular momentum.  This obviously implies that at large $\ell$ the pair of particles will become well-separated, although due to the warped AdS geometry, their separation or impact parameter $b$ is
\be
b \approx R_{AdS} \log \left( \frac{\ell}{\Delta_{\phi}} \right)
\ee
at large $\ell$.  So we need to study very large $\ell$ to create a large separation in AdS units.

In the absence of large $N$, we certainly cannot interpret the state $| \phi \rangle$ as a bulk particle, but we can still view it as some de-localized blob in AdS.   Without large $N$ we would also expect to lose the interpretation of operators in the $\phi(x) \phi(0)$ OPE as 2-particle states.  The results of the previous sections show that on the contrary, at large $\ell$ there are always operators $\CO_{\tau, \ell}$ in the OPE that we can interpret as creating `2-blob' states, where the blobs are orbiting each other at large separation in AdS.  The fact that we must always have infinite towers of operators in the OPE with twist $\tau = 2 \Delta_{\phi} + 2n  + \gamma(n, \ell)$ and $\gamma(n, \ell) \to 0$ as $\ell \to \infty$ shows that at large $\ell$, the interactions between these orbiting AdS blobs are shutting off.  In particular, let us assume that there is exactly one operator at each $n$ and $\ell$ and that the $\gamma(n, \ell) \to 0$ smoothly.  In this case we obtain a specific power-law dependence on $\ell$ that can be written as
\be
\gamma(n, \ell) = \frac{\gamma_n}{\ell^{\tau_m}} \propto \gamma_n  \exp \left[ - \tau_m \frac{b}{R_{AdS}}  \right],
\ee
so the interactions between the blobs are shutting off exponentially at large, superhorizon distances in AdS.  This is the sense in which our results prove superhorizon locality in the putative AdS dual of any $d > 2$ CFT.  

To emphasize the generality of this result, and the fact that $\phi$ really create `blobs', note that we can even apply our results to the scalar primary operators $\phi$ that create large black holes in AdS theories dual to CFTs with large $N$ and large 't Hooft coupling.  In that case, our results show that if the AdS black holes orbit each other with sufficiently large angular momentum, then their interactions become negligible.

\section{Discussion}
\label{sec:Discussion}

The recent revival of the conformal bootstrap has led to a great deal of progress, but perhaps the best is yet to come.    Thus far much of the work on the bootstrap has been numerical and has focused on questions of phenomenological interest, so further studies of superconformal theories \cite{Poland:2010wg}, AdS/CFT setups \cite{JP}, and even quantum gravity \cite{Analyticity} may yield important results.  Our results in this paper followed from a seemingly elementary consideration of how singularities in one channel of the conformal block expansion can be reproduced in the crossed channel, yet they have powerful implications for general CFTs.  

We have shown that the OPE of a scalar operator $\phi$ with itself has a universal leading behavior in the limit of large $\ell$ with fixed twist.  In particular, there always exist operators that we could call $[\phi \phi]_{n, \ell}$ at very large $\ell$ which have twist $2 \Delta_\phi + 2n + \gamma(n, \ell)$, with $\gamma(n, \ell) \to 0$ as $\ell \to \infty$.  This is directly analogous to the structure of `double-trace' operators in large $N$ theories, but it holds in any CFT.  Furthermore, we saw that with reasonable assumptions, we could make specific predictions for the fall-off of $\gamma(n, \ell)$ and of the related OPE coefficients.  We proved that all other contributions to the OPE must be sub-dominant at large $\ell$.  Our bootstrap methods apply in a simple way only to the OPE of scalar operators, but it seems very likely that equivalent results also hold for the OPE of higher-spin operators.  Perhaps in the future the results of \cite{Costa2011mg, Costa:2011mg} could be used to prove these statements.  Another interesting extension would involve studying further sub-leading corrections to the bootstrap as $u \to 0$; analyzing these corrections could lead to a more general proof of the Nachtmann theorem that does not rely on conformal symmetry breaking in the IR.

CFTs with dual AdS descriptions that are local at distances much smaller than the bulk curvature scale must have special features \cite{JP, AdSfromCFT}.  However, we have seen that in a certain technically precise sense, all $d > 2$ CFTs can be viewed as dual to AdS theories that are local at superhorizon distances.  The question of superhorizon AdS locality has often been discussed in the context of the holographic RG \cite{Susskind:1998dq, deBoer:1999xf, Balasubramanian:1999jd, Li:2000ec, Papadimitriou:2004rz, Heemskerk:2010hk, Faulkner:2010jy, vanRees:2011fr}, although the general success of this interesting approach has not been manifest.  It would be interesting if our results could be related to or shed light on the holographic RG.

Our arguments fail for CFTs in two dimensions.  In fact as we discussed in section \ref{sec:2d}, minimal models provide an immediate counter-example, as they have scalar operators of dimension $\Delta_{\phi}$ without corresponding operators of twist $\approx 2 \Delta_{\phi} + 2n$ at large $\ell$.  The reason is that in two dimensions, there is no separation between the dimension of the identity operator and the twists of conserved currents and the energy-momentum tensor.  One might try to interpret this in AdS as the statement that there is no clear separation between free propagation and interactions, perhaps due to the fact that gravitational interactions produce a deficit angle in three dimensions; it would be interesting to explore this issue further.

One inspiration for our approach was the structure of conformal blocks in Mellin space \cite{Mack, MackSummary, Analyticity, Unitarity, JoaoRegge}, where the blocks imitate the momentum-space partial waves of scattering amplitudes more transparently.  The leading behavior at small $u$ in position space translates into the presence of a leading pole in the Mellin amplitude which must be reproduced by an infinite sum over angular momenta in the crossed channel.  Through further work it should be possible to use our results to shed light on the convergence properties of the CFT bootstrap in Mellin space.  Our results seem to suggest that the sum of conformal blocks in Mellin space will only converge away from the region where the Mellin amplitude has poles.  A more precise version of this observation could be useful for further work using the CFT bootstrap, both analytically and numerically.

\section*{Acknowledgments}    

We are grateful to Sheer El-Showk, Ami Katz, Juan Maldacena, Miguel Paulos, Jo\~ao Penedones, Slava Rychkov, Alessandro Vichi, and Alexander Zhiboedov for discussions.  We would also like to thank the participants of the ``Back to the Bootstrap II" workshop for discussions and the Perimeter Institute for hospitality during the early stages of this work.  ALF and JK thank the GGI in Florence for hospitality while this work was completed; JK also thanks the University of Porto.  This material is based upon work supported in part by the National Science Foundation Grant No. 1066293. ALF was partially supported by ERC grant BSMOXFORD no. 228169. JK acknowledges support from the US DOE under contract no. DE-AC02-76SF00515.

\appendix
\newcommand\G{\Gamma}
\newcommand\R{\mathbb{R}}
\renewcommand\th{\theta}

\section{Properties of Conformal Blocks}
\label{app:LargeLBlocks}

Our arguments rely on a few key properties of crossed-channel conformal blocks in the small-$u$ (equivalently small-$z$) limit at asymptotically large values of $\ell$ and $\tau$.  In this Appendix we will establish these properties in general space-time dimensions.  We also establish some useful results that hold for general kinematics, including the positivity of the coefficients in the series expansion of the conformal blocks.

\subsection{Factorization at Large $\ell$ and Small $u$}

First, we would like to establish that in the large-$\ell$ and small-$u$ limits, the $\tau$- and $\ell$-dependence of the crossed-channel blocks factorizes.  More precisely, for the blocks in $d$ dimensions, we would like to show that
\be
\label{taulfactorization}
g_{\tau,\ell}^{(d)}(v,u) &\stackrel{\ell\gg 1}{\stackrel{u\ll 1}{=}}& k_{2\ell}(1-z) v^{\tau/2}F^{(d)}(\tau,v)\x(1 + O(1/\sqrt{\ell},\sqrt z)),
\ee
where
\be
k_{\beta}(x) \equiv x^{\beta/2} {}_2F_1\left(\beta/2,\beta/2,\beta,x\right),
\ee
and $F^{(d)}(\tau,v)$ is an analytic function that is regular and positive at $v = 0$.  In these expressions we are using the identifications $u=z\bar{z}$ and $v=(1-z)(1-\bar{z})$. The error term may depend arbitrarily on $\tau$ and $v$, but must have the indicated dependence on $\ell$ and $z \sim \frac{u}{1-v}$.
 
 \subsubsection{Factorization in $2$ and $4$ Dimensions}
Let us start by establishing this factorization in $d=2$. In this case the crossed channel blocks take the form
\be
g^{(2)}_{\tau,\ell}(v,u) &=& k_{2\ell+\tau}(1-z)k_{\tau}(1-\bar z)+k_{2\ell+\tau}(1-\bar z)k_{\tau}(1-z),
\ee
Since we will be in the regime with $(1-\bar{z}) < 1$, the second term is exponentially suppressed at large $\ell$ and we can ignore it.

Now, the hypergeometric function $k_{2\ell+\tau}(1-z)$ has the integral representation
\be
k_{2\ell+\tau}(1-z) &=& \frac{\G(2\ell+\tau)}{\G(\ell+\tau/2)^2}\int_0^1 \frac{dt}{t(1-t)}\p{\frac{(1-z)t(1-t)}{1-t(1-z)}}^{\tau/2}\p{\frac{(1-z)t(1-t)}{1-t(1-z)}}^{\ell},\nn\\
\ee
where we have factored the integrand into a $\tau$-dependent piece and an $\ell$-dependent piece.  When $\ell$ is large, the integrand is sharply peaked near the value $t_*=\frac{1-\sqrt{z}}{1-z}$, with a width that goes like $1/\sqrt{\ell}$.  Meanwhile, the $\tau$-dependent part of the integrand varies slowly over the peak, and thus contributes its value at $t_*$ (up to small corrections)
\be
\p{\frac{(1-z)t(1-t)}{1-t(1-z)}}^\tau 
&\sim & 
1+O(\sqrt z, 1/\sqrt\ell).
\ee
Plugging this in, and using Stirling's approximation for the $\G$-functions, we find
\be
k_{2\ell+\tau}(1-z) &=& 2^\tau k_{2\ell}(1-z)\x(1+O(\sqrt z, 1/\sqrt\ell)).
\ee
In the small $z$ limit, we have $1-\bar z = v + O(z)$, so that
\be\label{eq:Standard2dBlocks}
g^{(2)}_{\tau,\ell}(v,u) &=& k_{2\ell}(1-z)2^\tau k_{\tau}(v)\x(1 + O(\sqrt z, 1/\sqrt{\ell})).
\ee
This verifies Eq.~(\ref{taulfactorization}) in 2d with
\be
\label{eq:2dFfunction}
F^{(2)}(\tau,v) &\equiv& 2^\tau {}_2F_1\p{\frac{\tau}{2},\frac{\tau}{2},\tau,v}.
\ee

It is straightforward to repeat these steps in $d=4$. In this case, the crossed-channel blocks are given by
\be
g^{(4)}_{\tau,\ell}(v,u) &=& \frac{(1-z) (1-\bar z)}{\bar z- z}(k_{2\ell+\tau}(1-z)k_{\tau-2}(1-\bar z)-k_{2\ell+\tau}(1-\bar z)k_{\tau-2}(1-z)).
\ee
Again, one can neglect the second term in the large $\ell$ limit, and accounting for the pre-factor one straightforwardly finds
\be\label{eq:Standard4dBlocks}
g^{(4)}_{\tau,\ell}(v,u) &=& k_{2\ell}(1-z)\frac{2^\tau v}{1- v}k_{\tau-2}(v)\x(1 + O(1/\sqrt{\ell},\sqrt z)).
\ee
This verifies Eq.~(\ref{taulfactorization}) in 4d with
\be
\label{eq:4dFfunction}
F^{(4)}(\tau,v) &\equiv& \frac{2^\tau }{1-v}{}_2F_1\p{\frac{\tau}{2}-1,\frac{\tau}{2}-1,\tau-2,v}.
\ee

\subsubsection{Extension to Even Dimensions via a Recursion Relation for $F^{(d)}(\tau,v)$}
\label{sec:EvendRecursion}

One can easily extend this to all other even $d$ using a recursion relation relating the conformal blocks in $d$ dimensions to those in $(d+2)$ dimensions. Concretely, the blocks satisfy the relation~\cite{Dolan:2003hv}
\be
\left( \frac{z-\bar{z}}{z \bar{z}} \right)^2 g_{\De,\ell}^{(d+2)}(u,v) &=&  g_{\De-2,\ell+2}^{(d)}(u,v)  - 4\frac{(\ell-2)(d+\ell-1)}{(d+2\ell-2)(d+2\ell)} g_{\De-2,\ell}^{(d)}(u,v)  \\
&& -  4 \frac{(d-\De-1)(d-\De)}{(d-2\De)(d-2\De+2)} \left[ \frac{(\De+\ell)^2}{16(\De+\ell-1)(\De+\ell+1)} g_{\De,\ell+2}^{(d)}(u,v)  \right. \nn\\
&& \left. - \frac{(d+\ell-2)(d+\ell-1)(d+\ell-\De)^2}{4(d+2\ell-2)(d+2\ell)(d+\ell-\De-1)(d+\ell-\De+1)} g_{\De,\ell}^{(d)}(u,v)  \right]. \nn
\ee
In the large $\ell$ but fixed $\tau= \Delta-\ell$ limit, this recursion relation simplifies to
\be
\left( \frac{z-\bar{z}}{z \bar{z}} \right)^2 g_{\tau,\ell}^{(d+2)}(u,v)  &\approx&  g_{\tau-4,\ell+2}^{(d)} - g_{\tau-2,\ell}^{(d)}(u,v)  -  \frac{1}{16} g_{\tau-2,\ell+2}^{(d)}(u,v) \\
&&  + \frac{(d-\tau)^2}{16(d-\tau-1)(d-\tau+1)} g_{\tau,\ell}^{(d)}(u,v).  \nn
\ee
Inserting the factorized form Eq.~(\ref{taulfactorization}) and using the property $k_{2(\ell+2)}(1-z)\approx 2^4 k_{2\ell}(1-z)$, one finds that the function $F^{(d)}(\tau,v)$ should satisfy the recursion relation
\be
(1-v)^2 F^{(d+2)}(\tau,v) &\approx& 16  F^{(d)}(\tau-4,v) -  2 v F^{(d)}(\tau-2,v) \\
&&   + \frac{(d-\tau)^2}{16(d-\tau-1)(d-\tau+1)}  v^2 F^{(d)}(\tau,v)\nn .
\ee
This makes it trivial to generate the function $F^{(d)}(\tau,v)$ for any even $d$. Moreover one can verify that this relation holds between the functions $F^{(2)}(\tau,v)$ and $F^{(4)}(\tau,v)$ obtained above.

\subsubsection{Extension to Odd Dimensions}
Finding the function $F^{(d)}(\tau,v)$ in odd $d$ is less straightforward, but it is easy to see that the factorization property should continue to hold. One approach is to note that the conformal blocks are determined as the solutions to the Casimir equation \cite{Dolan:2000ut, Dolan:2003hv, Dolan:2011dv, DSDProjectors}
\be
\label{eq:CasimirDiffEq}
\mathcal{D} g_{\tau,\ell}^{(d)}(v,u) = \frac{1}{2} C^{(d)}_{\tau, \ell} g_{\tau,\ell}^{(d)}(v,u)
\ee
where $\mathcal{D}$ is the conformal Casimir as a differential operator
\be
\mathcal{D} &=&  (1-v-u) \partial_u u \partial_u + v \partial_v (2 v \partial_v - d)
- (1+v-u) (v \partial_v + u \partial_u)^2 
\ee
and $C^{(d)}_{\tau,\ell} = 2\ell(\ell+\tau-1) + \tau (\tau - d)$. One can immediately see that the $d$-dependence enters the Casimir at $O(1/\ell^2)$ at large $\ell$, so the dominant dependence is in a part of the differential operator which depends only on $v$.  Furthermore, it is also clear that we can define the blocks by integrating this differential equation, because from \cite{Dolan:2011dv} we have that in the small $v$ limit
\be
g^{(d)}_{\tau, \ell}(v, u) = v^{\frac{\tau}{2}} (1-u)^\ell {}_2F_1\left( \frac{\tau}{2} + \ell, \frac{\tau}{2} + \ell, \tau + 2 \ell , 1-u \right) + \CO(v)
\ee
This gives a boundary condition as $v \to 0$ for any value of $u$, so to obtain the conformal blocks at any $v$ we need only integrate equation (\ref{eq:CasimirDiffEq}) a small distance, over which the $\CO(u, \frac{1}{\ell})$ errors cannot accumulate.  Since the equation (\ref{eq:CasimirDiffEq}) is satisfied by the ansatz in equation (\ref{taulfactorization}) to leading order, the factorized ansatz suffices in all $d$.

\subsubsection{Further Approximations for the Function $k_{2\ell}(1-z)$}
\label{app:Bessel}

The function $k_{2\ell}(1-z) = (1-z)^{\ell} {}_2F_1(\ell,\ell,2\ell,1-z)$ can be approximated further depending on the relative size of $\ell$ and $z$. In the extreme asymptotic regime $\ell \gg 1/\sqrt{z}$ one can use the saddle point approximation
\be
\label{eq:FixedUStationaryPhase}
 \left(\frac{\Gamma(\ell)^2}{\Gamma(2\ell)}\right) {}_2F_1(\ell,\ell,2\ell,1-z) &=& \int_{0}^1 \frac{dt}{t(1-t)} \left( \frac{t(1-t)}{1-t (1-z)} \right)^\ell =  \int_{0}^1 \frac{dt}{t(1-t)} e^{\ell \ln \left( \frac{t(1-t)}{1-t (1-z)} \right)} \nn  \\
&\approx& \frac{1}{t(1-t)}  \sqrt{\frac{2\pi}{-\ell \frac{d^2}{dt^2} \ln \left( \frac{t(1-t)}{1-t (1-z)} \right) }} e^{\ell \ln \left( \frac{t(1-t)}{1-t (1-z)} \right)} \bigg|_{t = \frac{1-\sqrt{z}}{1-z}} \nn \\
&\approx& \sqrt{\frac{\pi}{\ell}}\frac{1}{\sqrt[4]{z} (1+\sqrt{z})^{2\ell-1}} \left( 1 + O \left( \frac{1}{ \ell\sqrt{z} }\right) \right) . 
\ee
The corrections can be easily obtained by expanding to the next order.  This approximation is best when we take $\ell \rightarrow \infty$ at a fixed value of $z$. In the extreme limit where $\sqrt{z} \ll 1$ but $\ell \sqrt{z} \gg 1$, one sees an exponential decay
\be
\label{eq:Exponentialdecay}
 \left(\frac{\Gamma(\ell)^2}{\Gamma(2\ell)}\right) {}_2F_1(\ell,\ell,2\ell,1-z) &\approx& \sqrt{\frac{\pi}{\ell}}\frac{e^{-2\ell\sqrt{z}}}{\sqrt[4]{z}} \left( 1 + O \left( \sqrt{z}, \frac{1}{ \ell\sqrt{z} }\right) \right) . 
\ee

Alternatively, we could consider the regime where we take $\ell \rightarrow \infty$ holding $y \equiv z \ell^2$ fixed. In this case, we can approximate
\be
\left( \frac{ \Gamma(\ell)^2}{\Gamma(2\ell)} \right) F(\ell,\ell,2\ell,1-z) &=& \int_0^1 \frac{dt}{t(1-t)} \left( \frac{t(1-t)}{1-t(1-x)} \right)^\ell \nn\\
&\stackrel{\ell \gg 1}{\approx} & \int_0^1 \frac{t^{\ell-1}}{1-t} e^{- \frac{t y}{(1-t) \ell} + O(\frac{1}{(1-t)^2 \ell^3})}.
\ee
The higher order terms in the exponent can be neglected at large $\ell$ with $y \lesssim O(1)$ held fixed.  If we define a new variable $s \equiv \frac{t}{1-t}$, we can rewrite the integral as
\be
\label{eq:FixedUApprox}
\left( \frac{ \Gamma(\ell)^2}{\Gamma(2\ell)} \right) F \left(\ell,\ell,2\ell,1-\frac{y}{\ell^2} \right) & \approx & \int_0^\infty \frac{ds}{s} e^{-\frac{s  y }{\ell} - \ell \frac{1}{s}} 
\nn \\ &=& 2 K_0(2 \sqrt{y} ) + O \left( \frac{1}{\ell} \right) ,
\ee
where $K_0$ is a modified Bessel function of the second kind.  We stress that this approximation breaks down when $y = z\ell^2 \gg 1$ (though it does correctly reproduce the exponential decay in Eq.~(\ref{eq:Exponentialdecay})), but provides a good description of the regime with $y \lesssim O(1)$.

\subsection{Positivity of Coefficients and Exponential Falloff at Large $\tau$}

In this section we will prove that in general $d$, the coefficients in the power series expansion of the conformal blocks
\be
g_{\tau,\ell}(z, \bar z) &=& \sum_{m,n\geq 0}^\infty a_{mn}z^{\tau/2+m}\bar z^{\tau/2+n} 
\ee
satisfy 
\be
a_{mn} > 0
\ee
for all $m, n$, where we recall that $u = z \bar z$ and $v = (1-z)(1-\bar z)$. One can verify the claim directly in the case of $d=2,4,6$ where explicit formulas are known.   Similar positivity results are fairly well-known \cite{Pappadopulo:2012jk} when the conformal blocks are expanded with $z = \bar z$, but as far as we are aware this more general result has not been discussed previously.  

This result will be useful because it implies that for real $z$ and $\bar z$, if we take $\bar z \to \bar z / \lambda $ with $\lambda > 1$ then we must have
\be
\label{eq:ExpDecreaseTau}
g_{\tau, \ell}(z, \bar z / \lambda )  <  \lambda^{-\frac{\tau}{2}} g_{\tau, \ell}(z, \bar z) .
\ee
This allows us to conclude that contributions from infinite sums over large $\tau$ are exponentially small for appropriate choices of kinematics, because for $0 < z , \bar z < 1$ we know that the conformal block expansion converges \cite{Pappadopulo:2012jk}.

The proof is simple, and uses the clever choice of coordinates in \cite{Pappadopulo:2012jk}, where we set $x_3 = e^{i \beta}$, $x_4 = -e^{i \beta}$, while $x_1 = r e^{i \alpha}$ and $x_2 = -r e^{i \alpha}$ with $\alpha, \beta, r$ all real and $0 \leq r < 1$.   In these coordinates we have $\rho \equiv r e^{i (\alpha - \beta)}$ given in terms of $z$ by
\be
\rho(z) = \frac{z}{(1 + \sqrt{1 -z})^2} ,
\ee
and we can define a quantity $\bar \rho( \bar z)$ similarly.  The formula can be inverted to give $z(\rho)$.  A crucial point is that when we expand $\rho(z)$ in a power series in $z$, all of the coefficients will be positive.  This implies that if $g_{\tau, \ell}(\rho, \bar \rho)$ has a power series expansion in $\rho, \bar \rho$ with positive coefficients, then the same property holds when it is viewed as a function of $z, \bar z$.  Note that 
\be
\frac{\rho^L + \bar \rho^L}{2} = r^\ell \cos(L (\alpha - \beta)) ,
\ee
so we can expand the conformal blocks as
\be
g_{\tau, \ell}(\rho, \bar \rho) = \sum_{s, L} c_{s, L} (\rho \bar \rho)^{\frac{\tau}{2} + s + \frac{L}{2}} \cos(L (\alpha - \beta)) .
\ee
If we can prove that $c_{s, L} > 0$ then we will have proven that $a_{mn} > 0$.  However, we can immediately see that
\be
\sum_{s, L} \int_0^{2 \pi} d\alpha \cos(m \alpha) \int_0^{2 \pi} d\beta \cos(m \beta)
 \langle \CO(e^{i \beta}) \CO(- e^{i \beta}) | \CO_{\tau + 2s, L} \rangle 
 \langle \CO_{\tau + 2s, L} | \CO(r e^{i \alpha }) \CO(- r e^{i \alpha } ) \rangle
\ee
is the norm of some definite linear combination of descendants of the primary $\CO_{\tau, \ell}$ whose conformal block we are considering.  Therefore this norm will be positive.  Applying this norm to the series expansion of the block above, we find
\be
\sum_s c_{s, L}\ \!  r^{\tau + 2s + L} > 0
\ee
for every $L$.  One can similarly project onto definite powers of $r$ by smearing $\CO(x_1) \CO(x_2)$ and $\CO(x_3) \CO(x_4)$ in the radial direction.    For example, we can promote $x_3 \to e^{\lambda_{34}} e^{i \beta}$ and $x_4 \to -e^{\lambda_{34}} e^{i \beta}$, and similarly take $x_1 = e^{-\lambda_{12}} e^{i \alpha}$ and $x_2 = -e^{-\lambda_{12}} e^{i \alpha}$, so that we have $r = e^{-\lambda_{12} - \lambda_{34}}$.  Then smearing the conformal block against wavefunctions such as $\cos(\tau_{12} \lambda_{12})$ and $\cos(\tau_{34} \lambda_{34})$ projects out definite values for $s$, and we have positivity for the case $\tau_{12} = \tau_{34}$.  

A more concise way of obtaining the same conclusion is to consider the $c_{s, L}$ as the norms of states on the subspaces of definite dimension and angular momentum on the $\rho$ circle.  In any case, we find that all $c_{s,L} > 0$, and so we conclude that $a_{mn} > 0$ as claimed.

\section{$\rho(\s)$ and its Crossing Equation}
\label{app:Rigorous}

In this Appendix, we present several details that were suppressed in section~\ref{sec:DblTrace} for the sake of readability.  Specifically, we will give a rigorous definition of the asymptotic density $\rho(\s)$, and a derivation of its crossing equation.  

\subsection{Existence of $\rho(\s)$}

Let us define a conformal block ``density" in $\tau$-space on the RHS of the crossing relation,
\be
D_{v,u}(\s) \equiv \p{\frac{u}{v}}^{\Delta_{\phi}}\sum_{\tau,\ell}\de(\tau-\s)P_{\tau,\ell} g_{\tau,\ell}(v,u).
\ee
One should always think of $D_{v,u}(\s)$ as being integrated against some function $f(\s)$.  More formally,  $D_{v,u}$ defines a linear functional given by the pairing
\be
\label{eq:linearfunctionaldefinition}
D_{v,u}[f] &\equiv& \p{\frac{u}{v}}^{\Delta_{\phi}}\sum_{\tau,\ell}f(\tau)P_{\tau,\ell} g_{\tau,\ell}(v,u).
\ee
Since the conformal block decomposition is absolutely convergent, $D_{v,u}$ is well-defined on any bounded function $f$.  Further, since the conformal blocks and coefficients $P_{\tau,\ell}$ are positive, $D_{v,u}$ is positive as well.  We will use $D_{v,u}$ as a tool for slicing up the RHS of the crossing relation into  different contributions at various twists.

Our goal is to characterize the limit $\lim_{u\to 0}D_{v,u}$ in terms of a density in twist-space at asymptotically large $\ell$.
To disentangle asymptotic behavior in $\ell$ from asymptotic behavior in $\tau$, it's extremely useful to restrict the twist $\tau$ to lie in some finite range, and study the contribution of only the operators in this range.  In terms of the functional $D_{v,u}$, this means we should consider its action on continuous functions with compact support, $f\in C_c(\R^+)$.\footnote{Of course, the twist is restricted by unitarity to lie above some minimum value depending on the spacetime dimension.  This is unimportant to the present analysis; for convenience, we will allow $f$ to be a function on all of $\R^+$.}  Note in particular that when $f$ has compact support, then there are only a finite number of nonzero terms at each $\ell$ on the RHS of Eq.~(\ref{eq:linearfunctionaldefinition}).  This will be important below.

Let us suppose $f\in C_c(\R^+)$, and consider the limit
\be
\lim_{u\to 0}D_{v,u}[f] &=& \lim_{u\to 0}\p{\frac{u}{v}}^{\Delta_\phi}\sum_{\tau,\ell}f(\tau)P_{\tau,\ell}g_{\tau,\ell}(v,u).
\ee
The limit of each individual term above vanishes, so the limit of the sum is unchanged if we discard any finite number of terms.  Since $f$ has compact support, we may restrict $\ell\geq L$ for any finite $L$ without changing the result, 
\be
\lim_{u\to 0}D_{v,u}[f] &=& \lim_{u\to 0}\p{\frac{u}{v}}^{\Delta_\phi}\sum_{\tau \ge 0, \ell\geq L}f(\tau)P_{\tau,\ell}g_{\tau,\ell}(v,u)\nn
\\
 &=& \lim_{z\to 0}\p{\frac{u}{v}}^{\Delta_\phi}\sum_{\tau \ge 0, \ell\geq L}f(\tau)P_{\tau,\ell}k_{2\ell}(1-z) v^{\tau/2}F^{(d)}(\tau,v)\x(1+O(1/\sqrt L,\sqrt z))\nn\\
  &=& \p{\lim_{z\to 0}\p{\frac{u}{v}}^{\Delta_\phi}\sum_{\tau, \ell\geq 0}f(\tau)P_{\tau,\ell}k_{2\ell}(1-z) v^{\tau/2}F^{(d)}(\tau,v)}\x(1+O(1/\sqrt L)).\nn\\
\ee
In the second line, we have substituted the small-$z$ and large-$\ell$ form of the crossed-channel blocks Eq.~(\ref{taulfactorization}).  In the third line, we have used the fact that the limit still vanishes termwise after this substitution.  The above equation holds for all finite $L$, so we may drop the error term to obtain
\be
\label{eq:limitintermsofDprime}
\lim_{u\to 0}D_{v,u}[f(\s)] &=& \lim_{z\to 0}\p{\frac{u}{v}}^{\Delta_\phi}\sum_{\tau,\ell\geq 0}f(\tau)P_{\tau,\ell}k_{2\ell}(1-z) v^{\tau/2}F^{(d)}(\tau,v)\nn\\
&=& (1-v)^{\De_\phi} D'\left[v^{\s/2-\De_\phi}F^{(d)}(\s,v)f(\s)\right],
\ee
where we have used $u\approx z(1-v)$ and defined the functional
\be
D'[f] &\equiv& \lim_{z\to 0} z^{\Delta_\phi}\sum_{\tau,\ell} P_{\tau,\ell}f(\tau) k_{2\ell}(1-z).
\ee

$D'$ is a positive linear functional on $C_c(\R)$.  By the Reisz Representation Theorem (RRT), we can write it as the integral of a density in $\tau$-space.  Specifically, the theorem states that there exists a unique regular Borel measure, which we denote by $\rho(\s)d\s$, such that
\be
\label{eq:definitionofrho}
D'[f] &=& \int_0^\oo f(\s)\rho(\s)d\s
\ee
for all $f$ in $C_c(\R)$.\footnote{The density $\rho(\s)$ can be reconstructed by considering the value of $D'[f]$ on functions with support in very small neighborhoods.  The non-trivial content of this theorem is that positivity and linearity imply that the density so-constructed is unique, and gives the correct value of the functional on any set with compact support.}

In summary, we may formally define a density $\rho(\s)$ by
\be
\rho(\s) &=& \lim_{z\to 0} z^{\De_\phi}\sum_{\tau,\ell} P_{\tau,\ell}\de(\tau-\s)k_{2\ell}(1-z).
\ee
When integrating $\rho(\s)$ against some function $f$, one might worry about switching the order of the limit and the integration.  We can interpret the RRT as saying that switching the limit and integration is justified  whenever $f$ has compact support.

\subsection{Crossing Symmetry for $\rho(\s)$}

Crossing symmetry states that
\be
\label{eq:crossingsymmetryintermsofDprime}
1 &=& \lim_{u\to 0} D_{v,u}[1]
\ \ =\ \   (1-v)^{\De_\phi} D'\left[v^{\s/2-\De_\phi}F^{(d)}(\s,v)\right] .
\ee
We can't immediately write this in terms of the density $\rho(\s)$ because a priori Eq.~(\ref{eq:definitionofrho}) only holds for functions $f$ with compact support.  This is the order of limits and integration issue mentioned in section~\ref{sec:DblTrace}.
For example, one might worry that the limit $\lim_{u\to 0} D_{v,u}[1]$ is dominated by operators with $\tau\sim\ell\sim 1/u$.  Then $D'[f]$ would vanish on functions with compact support (and $\rho(\s)$ would be identically zero), while Eq.~(\ref{eq:crossingsymmetryintermsofDprime}) might still hold.  

We can show that this {\it does not} happen by using the fact that the blocks die exponentially at large $\tau$.  Specifically, let us study the contribution of conformal blocks with twist less than some large $\tau_*$.  Since the function $\th(\tau_*-\s)$ has compact support, Eqs.~(\ref{eq:limitintermsofDprime}) and~(\ref{eq:definitionofrho}) imply\footnote{Strictly speaking, we should use a slightly smoothed $\th$-function, since $\th$ itself is not continuous.  This subtlety will not be important in the analysis.}
\be
(1-v)^{\De_\phi}\int_0^{\tau_*}v^{\s/2-\De_\phi}F^{(d)}(\s,v)\rho(\s)d\s 
&=&
\lim_{u\to 0} D_{v,u}[\theta(\tau_*-\s)] ,\nn\\
&=& \lim_{u\to 0}\p{\frac{u}{v}}^{\Delta_\phi}\sum_{\tau\leq\tau_*}P_{\tau,\ell}g_{\tau,\ell}(v,u).
\label{eq:compactsupportstatement}
\ee

Our goal is to bound the discrepancy between this quantity and the full conformal block expansion.
Let us choose some constant $\lambda$ satisfying $1<\lambda<\frac{1}{1-\bar z}$. Eq.~(\ref{eq:ExpDecreaseTau}) then implies
\be
\lim_{u\to 0}\p{\frac{u}{v}}^{\Delta_\phi}\sum_{\tau>\tau_*}P_{\tau,\ell}g_{\tau,\ell}(v,u)
&\leq&
\lambda^{-\tau_*}\lim_{u\to 0}\p{\frac{u}{v}}^{\Delta_\phi}\sum_{\tau>\tau_*}P_{\tau,\ell}g_{\tau,\ell}(v,u)|_{\bar z\to\bar z'} ,\nn\\
&\leq&
\lambda^{-\tau_*}\lim_{u\to 0}\p{\frac{u}{v}}^{\Delta_\phi}\sum_{\tau,\ell}P_{\tau,\ell}g_{\tau,\ell}(v,u)|_{\bar z\to\bar z'} ,\nn\\
&=&\lambda^{-\tau_*},
\label{eq:largetwistbound}
\ee
where in the second line we've used positivity of each term in the conformal block expansion to enlarge the sum to all operators, and in the last line we've used crossing symmetry at the point $(z,\bar z')$.

Combining Eqs.~(\ref{eq:compactsupportstatement}) and~(\ref{eq:largetwistbound}), we find
\be
 1+O(\lambda^{-\tau_*}) &=& (1-v)^{\De_\phi}\int_0^{\tau_*}v^{\s/2-\De_\phi}F^{(d)}(\s,v)\rho(\s)d\s
,
\ee
so that we can safely extend the region of integration out to infinity and conclude
\be
1 &=& (1-v)^{\De_\phi}\int_0^{\oo}v^{\s/2-\De_\phi}F^{(d)}(\s,v)\rho(\s)d\s.
\ee
This is Eq.~(\ref{eq:integralcrossingrelation}) in the text.

\subsection{Bounds on OPE Coefficient Densities}
\label{app:BoundsF}

In this Appendix, we will prove a bound on the asymptotic behavior of the integral over $f(\ell)$, which is defined in Eq.~(\ref{eq:fLDefinition}).  Specifically, we will show that given a function $\CL(x)$ with the representation 
\be
\CL(x) &=& \int_0^\infty d\beta f(\beta) k_\beta(1-x^2)
\ee
that behaves like $x^{-a}$ at small $x$, then there exist numbers $A_L, A_U$ such that the integrated value
\be
F(B) &\equiv& \int_0^B d\beta \bar{f}(\beta) , \qquad  \bar{f}(\beta) \equiv  \frac{\Gamma(\beta)}{\Gamma^2(\beta/2)} f(\beta),
\ee
is bounded at large $B$ by
\be
A_U B^a > F(B) > A_L \frac{B^a}{\log(B)}.
\ee
We will begin with the upper bound.  First, we define 
\be
h(\beta,x) &\equiv & \frac{\Gamma^2(\beta/2)}{\Gamma(\beta)} k_\beta(1-x^2),
\ee
which is a positive, decreasing function of $\beta$ at fixed $x$.  Since $\bar{f}(\beta) h(\beta,x)$ is non-negative, we therefore have for any $B$ that
\be
\CL(x) \ge \int_0^B d\beta \bar{f}(\beta) h(\beta,x) \ge h(B,x) F(B).
\ee
Fixing $\lambda$ to a positive real number (its value is not important) and taking $x=\lambda/B$, we therefore find the bound
\be
F(B) \le \frac{B^a}{\lambda^a h(B,\frac{\lambda}{B})}
\ee
at sufficiently large $B$.    Furthermore, as shown in Appendix \ref{app:Bessel}, in the limit of large $B$ with $\lambda$ fixed, $h(B,\frac{\lambda}{B})$ approaches a finite value independent of $B$:
\be
\lim_{B\rightarrow \infty} h(B,\frac{\lambda}{B}) = 2 K_0(\lambda),
\ee
where $K_0(\lambda)$ is a modified Bessel function.  Thus, we have proven an upper bound
\be
\lim_{B\rightarrow \infty} B^{-a} F(B) \le \frac{1}{\lambda^a 2 K_0(\lambda)}.
\ee
One can try to improve this bound by maximizing $\lambda^a K_0(\lambda)$ as a function of $\lambda$, but any value of $\lambda$ proves the existence of some $A_U$.

Now, let us turn to proving the lower bound.  It will be convenient to define the function $\tilde{h}(\beta,u)$ by
\be
\tilde{h}(\beta, x) = -\frac{\partial}{\partial \beta} h(\beta, x) ,
\ee
which is positive since $h(\beta,x)$ is decreasing as a function of $\beta$. Then, we can write
\be
\CL(x) &=& \int_0^\infty dB F(B) \tilde{h}(B,x).
\ee
The limit $\lim_{x\rightarrow 0} x^a \CL(x)$ is unaffected by adding a fixed lower-bound on the $d\beta$ integration, so we can define the function
\be
\tilde{\CL}(x) &\equiv& \int_{\beta_0}^\infty dB F(B) \tilde{h}(B,x),
\ee
which also approaches $x^{-a}$ at small $x$.  Furthermore, $F(B)$ is an integral over non-negative terms, so $F(B)\le F(B')$ for $B\le B'$. 
Thus, we have
\be
\tilde{\CL}(x) &=& \int_{\beta_0}^M dB F(B) \tilde{h}(B,x) + \int_M^\infty dB F(B) \tilde{h}(B,x) , \nn\\
  &\le& F(M) \left( h(\beta_0,x) - h(M,x) \right) + A_U \int_M^\infty dB B^{2a} \tilde{h}(B,x) , \nn\\
   &\le& F(M) h(\beta_0, x) + A_U \int_M^\infty dB B^{2a} \tilde{h}(B,x) .
  \ee
Let us take $x=\lambda/M$ such that $1<\lambda <M$.  The second integral on the right-hand-side above is just an integral over known functions and by using the large $B$, large $x B$ approximation to $\tilde{h}(B,x)$, it can be seen to vanish like $A_2 e^{-\lambda}$ at large $\lambda$ and $M$ for some $A_2$.  Finally, the small $x$ behavior of $h(\beta_0, x)$ is proportional to $\log(1/x)$, so we have $h(\beta_0, x) < A_3 \log(1/x)$ for some $A_3$, and thus
  \be
  \tilde{\CL}(\lambda/M) \le F(M) A_3 \log(M/\lambda) + A_U A_2 e^{-\lambda}.
  \ee
  By making $\lambda$ large, we can make the last term as small as we like, so we have
  \be
 \lim_{M\rightarrow \infty} \log(M) M^{-a} F(M) > \frac{1}{A_3 \lambda^a },
  \ee
 which proves the lower bound.  

Finally, note that if we knew that the bounds could be improved such that the upper and lower bound were the same ($F(B)\sim A_M B^a$ for some constant $A_M$ that depended only on $a$), then we could calculate $A_M$ using the special case $\bar{f}(\beta) \propto \beta^{a-1}$ to be
\be
F(B) \to \frac{\int_0^B d\beta \beta^{a-1} }{ \lim_{x \rightarrow 0} x^a \int_0^\infty d\beta \beta^{a-1} h(\beta, x)} = \frac{\frac{1}{a}B^a}{\int_0^\infty d\lambda \lambda^{a-1} 2 K_0(\lambda)} = \left( \frac{B}{2} \right) ^a \frac{1}{\frac{a}{2}\Gamma^2(\frac{a}{2} )}.
\ee

\section{Generalization to Distinct Operators $\phi_1$ and $\phi_2$}
\label{app:DistinctOperators}

For simplicity we have presented the argument in the body of the draft for a single operator $\phi$.  However, our results extend to distinct operators $\phi_1$ and $\phi_2$, as might be expected from AdS intuition.  Specifically, by studying the correlator
\be
\langle \phi_1(x_1) \phi_1(x_2) \phi_2 (x_3)  \phi_2(x_4) \rangle
\ee
and using the crossing relation in the $12 \to 34$ and $13 \to 24$ channels, we can prove that there must exist an infinite sequence of operators with twists $\Delta_1 + \Delta_2 + 2n + \gamma(n, \ell)$ at large $\ell$.  If we have a unique such operator at each $\ell$, we can also show that
\be
\gamma(n, \ell) = \frac{\gamma_n}{\ell^{\tau_m}}  \ \ \ \mathrm{and} \ \ \ P_{n, \ell} = \frac{c_n}{\ell^{\tau_m}}
\ee
as we found for the case $\phi_1 = \phi_2$ in the body of the paper, where $\tau_m$ is the minimal non-vanishing twist appearing in the OPE of $\phi_1$ with itself.  The main complication compared to the case of identical operators is that the conformal blocks now depend on $2 \Delta_{12} = \Delta_1 - \Delta_2$, although not in any way that qualitatively affects the results in the $u \ll v < 1$ limit.  Note also that in the $12 \to 34$ channel we cannot assume that the conformal block coefficients are positive, although this will not be necessary to obtain a proof.

Note that for example in mean field theory we would have the single term
\be
\langle \phi_1(x_1) \phi_1(x_2)  \phi_2(x_3) \phi_2(x_4) \rangle &=& \frac{1}{(x_{12}^2)^{\Delta_1} (x_{34}^2)^{\Delta_2}} 
\ee
and we can factor this dependence out of all conformal blocks.  Then the crossing relation takes the form
\be
u^{-\frac{\Delta_1 + \Delta_2}{2}} + u^{-\frac{\Delta_1 + \Delta_2}{2}} \sum_{\tau, \ell} P_{\tau, \ell}^{11,22} g_{ \tau, \ell}^{11,22} (u,v) =  v^{-\frac{\Delta_1 + \Delta_2}{2}} \sum_{\tau, \ell} P_{\tau, \ell}^{12,12} g_{ \tau, \ell}^{12,12} (v,u) 
\ee
where we have labeled the conformal blocks and their coefficients according to the configuration of operators.  Note that on the left-hand side the blocks $g_{ \tau, \ell}^{11,22} (u,v)$ have leading terms at small $u$ proportional to  $u^{\frac{\tau}{2}}$, and for $d>2$ this is isolated from the identity.  In fact, $g^{11,22}_{\tau, \ell} = g_{\tau, \ell}$ that we used in the case of identical operators \cite{Dolan:2011dv}, as the difference $\Delta_1 \neq \Delta_2$ is irrelevant in this channel.  On the right-hand side the blocks $g_{ \tau, \ell}^{12,12} (v,u)$ again begin as $v^{\frac{\tau}{2}}$ at small $v$ \cite{Dolan:2011dv}.

Now we can proceed as in the body of the draft, expanding at small $u$ and keeping only the identity and the first sub-leading term on the left-hand side.  This gives an approximate crossing relation
\be
u^{-\frac{\Delta_1 + \Delta_2}{2}} + u^{\frac{\tau_m - \Delta_1 - \Delta_2}{2}} P_{m}^{11,22}  f_{ \tau_m, \ell_m} (u,v)
\approx
\sum_{\tau, \ell} P^{12,12}_{\tau, \ell} v^{\frac{\tau - \Delta_1 - \Delta_2 }{2}} f_{ \tau, \ell}^{12,12} (v,u) 
\ee
where we have separated out the leading $u$ dependence from the blocks to make the behavior clear.  This means that the sub-leading term on the left-hand side has a convergent series expansion in integer powers $v^k$ and $v^k \log v$, which we can use as in the case of identical operators to conclude that on the right-hand side we need contributions from towers of operators with twist approaching $\Delta_1 + \Delta_2 + 2 n$.

As we considered in Appendix \ref{app:LargeLBlocks}, the conformal blocks $g_{ \tau, \ell}^{12,12} $ can be determined as the solutions to the generalized Casimir equation \cite{Dolan:2000ut, Dolan:2003hv, Dolan:2011dv, DSDProjectors}
\be
\mathcal{D} g(v,u) = \frac{1}{2} C^{(d)}_{\Delta, \ell} g(v,u)
\ee
where $\mathcal{D}$ is the conformal Casimir as a differential operator that for $\Delta_{12} \neq 0$ is
\be
\mathcal{D} &=&  (1-v-u) \partial_u (u \partial_u - \Delta_{12}) + v \partial_v (2 v \partial_v - d)
\nonumber \\
&& - (1+v-u) (v \partial_v + u \partial_u - \Delta_{12})(v \partial_v + u \partial_u) ,
\ee
and $C^{(d)}_{\tau,\ell} = 2\ell(\ell+\tau-1) + \tau (\tau - d)$ as before. This is sufficient to conclude that
\be
g_{\tau, \ell}^{12,12} (v,u) \approx k_{2\ell}(1-z) v^{\tau/2}F^{(d)}(\tau,\Delta_{12}, v)\x(1 + O(1/\sqrt{\ell},\sqrt z)) ,
\ee
so we obtain the same leading order behavior at large $\ell$ as in the case of identical operators, and $F$ has some convergent series expansion in integer powers of $v$.  This means that the arguments from the body of the text can now be followed as before with $\phi_1$ and $\phi_2$, leading to analogous conclusions.

\section{Comparison With Gravity Results}
\label{app:Joao}

The calculation of anomalous dimensions of double-trace operators starting with a perturbative AdS description is usually simplest for the lowest-spin states \cite{Katz, Fitzpatrick:2011hh}, and becomes increasingly more involved for higher spins.  However, a significant simplification occurs in the limit of very large spin, which is described by the Eikonal limit of 2-to-2 scattering in AdS space, as was worked out in detail in \cite{ourEikonal, ourCPW}.  There, it was shown that the anomalous dimensions could be calculated by treating one of the two particles in AdS as a classical ``shock wave''  that sourced the deformation of the geometry in which the other particle traveled.  
The result of their calculations was that in the limit of large spin with  $\tau$ fixed and large, the anomalous dimensions due to semi-classical gravitational interactions behave as 
\be
 \gamma(\tau,\ell) \stackrel{\ell\gg 1}{\approx} - 2^{5-\Delta_m - j_m} \pi^{1-\frac{d}{2}} G \frac{\Gamma(\Delta_m-1)}{\Gamma(\Delta_m-\frac{d}{2} +1)} \frac{\tau^{\Delta_m + j_m-2}}{\ell^{\tau_m}}.
 \label{eq:joao}
\ee
The gravity case takes $\Delta_m=d$, $j_m=2$, and the coupling $G$ defined in \cite{ourEikonal, ourCPW} determines the strength of gravitational interactions, i.e. $G\propto G_N$.  This can then be compared to our formula (\ref{eq:n0matching})
\be
\gamma_0 = -P_{m} \frac{2\Gamma(\Delta_{\phi})^2 \Gamma(\tau_m+2 \ell_m)}{ \Gamma(\Delta_{\phi}-\frac{\tau_m}{2})^2\Gamma(\frac{\tau_m}{2}+\ell_m)^2} 
\ee 
combined with the specific gravity prediction for $P_m$.  Taking the specific values of $\tau_m=d-2$ and $\ell_m=2$, this reduces  to
\be
\gamma_0&=& \stackrel{\Delta_{\phi}\gg 1}{\approx} -P_m \Delta_{\phi}^{d-2}   \frac{ 2 \Gamma(d+2) }{\Gamma^2(\frac{d}{2} +1)}.
   \ee
   In $d=4$ the OPE coefficient of the stress tensor is $P_m = \frac{\Delta_{\phi}^2}{360 c}$,
   and therefore formula (\ref{eq:n0matching}) gives 
    \be
   \gamma(2\Delta_{\phi}, \ell) &\approx& -\frac{1}{6c}  \frac{\Delta_{\phi}^4}{\ell^2}.
   \ee
This is to be compared with (\ref{eq:joao}) for the gravity case with $\tau = 2 \Delta_{\phi}$ and $d=4$, which gives
   \be
   \gamma(2\Delta_{\phi}, \ell) &\approx& -\frac{8 G }{\pi}  \frac{\Delta_{\phi}^4}{\ell^2}.
   \ee
 The anomaly coefficient $c$ is proportional to $1/G_N$ in AdS gravity theories, so our results clearly agree parametrically with the perturbative AdS computations, with exact agreement for $c =  \frac{\pi}{48 G}$.
   
\bibliographystyle{utphys}
\bibliography{AnalyticBootstrapBib}

 \end{document}